\newif\ifsingle

\ifsingle
\documentclass[12pt,draftclsnofoot, onecolumn]{IEEEtran}		
\else		
\documentclass[10pt,final, twocolumn]{IEEEtran}
\fi

\usepackage{times}
\usepackage{amsmath,dsfont}
\usepackage{amssymb,amsthm}
\usepackage{epsfig,verbatim}
\usepackage{subfigure}
\usepackage{setspace}
\usepackage{color}
\usepackage{cite}
\usepackage{epstopdf}
\usepackage{graphics}
\usepackage{accents}
\usepackage{acronym}
\usepackage[bookmarks,colorlinks]{hyperref}
\usepackage{booktabs}
\usepackage{mathtools}
\usepackage{algorithm}
\usepackage{algorithmic}

\newtheorem{definition}{Definition}
\newtheorem{theorem}{Theorem}
\newtheorem{corollary}{Corollary}
\newtheorem{proposition}{Proposition}
\newtheorem{lemma}{Lemma}

\definecolor{NewColor}{rgb}{0,0,0}

\newcommand{\myVec}[1]{{\bf{#1}}}
\newcommand{\myMat}[1]{{\boldsymbol{#1}}}
\newcommand{\mySet}[1]{\mathcal{#1}}
\newcommand{\E}{\mathds{E}}		 			
\newcommand{\myI}{{\myMat{I}}}			 		
\newcommand{\myX}{{\myVec{x}}}			 		
\newcommand{\myZ}{{\myVec{z}}}			 		

\newcommand{\Rs}{R_{\rm{s}}}

\newcommand{\CovMat}[1]{\myMat{\Sigma}_{#1}}			
\newcommand{\PSDMat}[1]{\myMat{S}_{#1}}
\newcommand{\HmatFD}{{\myMat{\Gamma}}}			 			
\newcommand{\GmatFD}{{\myMat{\Sigma}}}			 			

\newcommand{\myW}{{\myVec{w}}}			 		
\newcommand{\myY}{{\myVec{y}}}			 		
\newcommand{\Mem}{m_h}
\newcommand{\MemG}{m_g}
\newcommand{\Mopt}{^{\rm OM}}						
\newcommand{\opt}{^{\rm OD}}						
\newcommand{\am}{^{\rm AM}}							
\newcommand{\ame}{^{\rm AM2}}						
\newcommand{\Nint}{B}
\newcommand{\SigW}{\sigma_W^2}						
\newcommand{\Nantennas}{N}						
\newcommand{\Nusers}{U}							
\newcommand{\Ndmas}{K}							
\newcommand{\NGmat}{{\rm rank}(\EqGMat)}
\newcommand{\Nelements}{L}						
\newcommand{\CovW}[1][]{\myMat{C}_{W_{#1}}}			
\newcommand{\CovWBar}[1][]{\bar{\myMat{C}}_{W_{#1}}}			
\newcommand{\Gmat}{{\myMat{G}}}			 			
\newcommand{\Hmat}{{\myMat{H}}}			 			
\newcommand{\Hfilt}{h}			 			
\newcommand{\myWeights}{{\myMat{Q}}}			 	
\newcommand{\myWeightSet}{{\mySet{Q}}}			 	
\newcommand{\myWeightScalar}{q}	
\newcommand{\EqGMatFD}{\tilde{\GmatFD}}
\newcommand{\EqGMat}{\tilde{\Gmat}}
\newcommand{\BlkDiag}{{\rm BlkDiag}}

\acrodef{rv}[RV]{random variable}
\acrodef{mse}[MSE]{mean-squared error}
\acrodef{mmse}[MMSE]{minimum mean-squared error}
\acrodef{dma}[DMA]{dynamic metasurface antenna}
\acrodef{cir}[CIR]{channel impulse response}
\acrodef{csi}[CSI]{channel state information}
\acrodef{sar}[SAR]{synthetic-aperture radar}

\acrodef{adc}[ADC]{analog-to-digital convertor}
\acrodef{dtft}[DTFT]{discrete-time Fourier transform}
\acrodef{psd}[PSD]{power spectral density}
\acrodef{dt}[DT]{discrete-time}
\acrodef{ct}[CT]{continuous-time}
\acrodef{awgn}[AWGN]{additive white Gaussian noise}
\acrodef{wss}[WSS]{wide-sense stationary} 

\acrodef{ofdm}[OFDM]{orthogonal frequency division multiplexing}
\acrodef{mimo}[MIMO]{multiple-input multiple-output}
\acrodef{map}[MAP]{maximum a-posteriori probability} 
\acrodef{isi}[ISI]{intersymbol interference}
\acrodef{snr}[SNR]{signal-to-noise ratio}
\acrodef{pc}[PC]{proper-complex}
\acrodef{pdf}[PDF]{probability density function}
\acrodef{bs}[BS]{base station} 
\acrodef{mac}[MAC]{multiple access channel}
\acrodef{tdd}[TDD]{time-division duplex}
\acrodef{ut}[UT]{user terminal}
\acrodef{ps}[PS]{pilot sequence}
\acrodef{rf}[RF]{radio frequency}
\acrodef{se}[SE]{spectral efficiency}
\acrodef{sinr}[SINR]{signal-to-interference-and-noise ratio}
\acrodef{svd}[SVD]{singular valued decomposition}

\long\def\symbolfootnote[#1]#2{\begingroup\def\thefootnote{\fnsymbol{footnote}}\footnote[#1]{#2}\endgroup}

\ifsingle
\newcommand{\figWidth}{0.65\columnwidth}

\else
\newcommand{\figWidth}{\columnwidth}

\fi 

 \IEEEoverridecommandlockouts
\title{Dynamic Metasurface Antennas for Uplink Massive MIMO Systems
}

\vspace{-0.75cm}

\author{
	\IEEEauthorblockN{Nir Shlezinger, Or Dicker, Yonina C. Eldar, Insang Yoo, Mohammadreza F. Imani, and  David R. Smith\\
	}	
	\thanks{Parts of this work were accepted for presentation in the 2019 IEEE International Conference on Acoustics, Speech, and Signal Processing (ICASSP), Brighton, UK.
	}
	\thanks{This project has received funding from the Air Force Office of Scientific Research under grants No. FA9550-18-1-0187 and FA9550-18-1-0208.
}
	\thanks{N. Shlezinger and Y. C. Eldar are with the faculty of Mathematics and Computer Science, Weizmann Institute of Science, Rehovot, Israel (e-mail: nirshlezinger1@gmail.com; yonina@weizmann.ac.il). 	
}
	\thanks{O. Dicker is with the department of Electrical Engineering, Technion, Haifa, Israel (e-mail: or.dicker@gmail.com ). 	
	}
	\thanks{I. Yoo, M. F. Imani and D. R. Smith are with the department of Electrical and Computer Engineering, Duke University, Durham, NC (e-mail: insang.yoo@duke.edu, mohamad.imani@gmail.com; drsmith@duke.edu).  	
	}
	
	\vspace{-1.0cm}
	
}

\vspace{-0.75cm}

\begin{document}

\maketitle
\pagestyle{plain}
\thispagestyle{plain}
\begin{abstract}
	Massive multiple-input multiple-output (MIMO) communications  are the focus of considerable interest in recent years. While the theoretical gains of massive MIMO have been established, implementing MIMO systems with large-scale antenna arrays in practice is challenging. Among the practical challenges associated with massive MIMO systems are increased cost, power consumption, and physical size. In this work we study the implementation of massive MIMO antenna arrays using dynamic metasurface antennas (DMAs), an emerging technology which inherently handles the aforementioned challenges. Specifically, DMAs realize large-scale planar antenna arrays, and can adaptively incorporate signal processing methods such as compression and analog combining in the physical antenna structure, thus  reducing the cost and power consumption. 
	We first propose a mathematical model for massive MIMO systems with DMAs and discuss their constraints compared to ideal antenna arrays. Then, we characterize the fundamental limits of uplink communications with the resulting systems, and propose two algorithms for designing practical DMAs for approaching these limits. Our numerical results indicate that the proposed approaches result in practical massive MIMO systems whose performance is comparable to that achievable with ideal antenna arrays.  
	
	{\textbf{\textit{Index terms---}} Massive MIMO, metasurfaces, antenna design.}
\end{abstract}

\vspace{-0.2cm}
\section{Introduction}
Future wireless systems are required to support an increasing number of end-users with growing throughput demands. 
Recent years have witnessed a rising interest in massive \ac{mimo} systems, in which the \ac{bs} is equipped with a large antenna array, as a method for meeting these demands and increasing the \ac{se}. 
In particular, it was shown that, when a sufficiently large number of antennas are utilized, the throughput can be increased in a manner which is scalable with the number of \ac{bs} antennas~\cite{Marzetta:15}. 

The theoretical benefits of massive \ac{mimo} systems in terms of \ac{se} are well-established \cite{Marzetta:10,Hoydis:13, Shlezinger:17}. However,  implementing a massive \ac{mimo} \ac{bs} equipped with a standard antenna array, capable of achieving these benefits, is still a very challenging task. In particular, some of the difficulties which arise when realizing large-scale  antenna arrays include high cost \cite{AlKhateeb:14,Rial:16}, increased power consumption \cite{Mo:17}, and constrained physical size and shape \cite{Akyildiz:16,Hoeher:17}. Several signal processing methods have been studied, aimed at tackling these difficulties. The proposed approaches include introducing analog combining to reduce the size and cost of the system \cite{AlKhateeb:14,Stein:17}; implementing low-resolution quantization and/or antenna selection to mitigate the power consumption \cite{Mo:17,Li:17,Choi:16,Choi:17,Choi:18,Zhang:16};  and utilizing efficient power amplifiers operating at reduced peak-to-average-power ratio  \cite{Mohammed:13,Studer:13}. Nonetheless, all these approaches assume a fixed optimal antenna array, and attempt to tackle the  difficulties which arise from this antenna array architecture from a signal processing perspective.

In parallel to the ongoing efforts to make massive \ac{mimo} feasible using signal processing techniques, a large body of research has focused on designing practical antenna arrays for massive \ac{mimo} systems \cite{Akyildiz:16, Ghosh:14, Ouedraogo:12, Hoeher:17}. An emerging technology for realizing large-scale antenna arrays of small physical size uses metamaterial radiators instead of conventional ones. Metamaterial antennas consist of array of subwavelength metamaterial radiators, excited by a waveguide or cavity \cite{Smith:17}. While the resulting antenna arrays typically exhibit mutual coupling and frequency selectivity, they offer comparable beam tailoring capability from a simplified hardware, which uses much less power and costs less than antenna arrays based on standard antenna arrays \cite{Mikala:15,Mikala:16}. {Furthermore, a large number of tunable metamaterial antenna elements can be packed in the same physical area, \cite{Akyildiz:16}, and  metasurfaces can implement planar antennas, making it an appealing technology for supporting the increased \ac{bs} deployment of 5G wireless networks \cite[Sec. II]{Andrews:14}.} 
%
Most previous works on metamaterial antennas for \ac{mimo} communications focus on designing the physical antenna structure and metamaterial substrate to satisfy desired requirements, such as gain, bandwidth, efficiency, and level of mutual correlation \cite{Akyildiz:16, Ghosh:14, Mookiah:09, Ouedraogo:12}. 
Consequently, the resulting antenna structure is fixed and independent of the processing which the transmitted and received signals undergo. An alternative application for metasurfaces as reflecting elements instead of as transmit or receive antennas, was proposed in \cite{Huang:18,Kaina:14,Hougne:18} as a scheme for improving energy efficiency in wireless communication networks.

Recently, \acp{dma} have been proposed as a method for electrically tuning the physical characteristics of metamaterial antennas \cite{Hunt:13, Smith:17,Sleasman:16}.  
\acp{dma} inherently implement signal processing techniques such as beamforming, analog combining, compression, and antenna selection, without additional hardware. 
By introducing simple solid-state switchable components into each metamaterial element and addressing them independently, these capabilities can become reconfigurable; i.e. they can adapt to the task at hand or changes in the environment. 
The application of \acp{dma} was shown to yield simple, fast, planar, and low-power systems for  microwave imaging \cite{Diebold:18, Sleasman:16b,Sleasman:17b}, radar systems  \cite{Sleasman:17,Boyarsky:17,Watts:17}, and satellite communications \cite{Johnson:15}. 
More recently, using cavity-backed \acp{dma} as a novel means to generate desired patterns to enhance capacity of \ac{mimo} communications in a clustered environment has been proposed and demonstrated in numerical simulations \cite{Yoo:18}.
 Nonetheless, despite the potential of \acp{dma} in combining signal processing and antenna design, their application  for realizing massive \ac{mimo} systems has not yet been studied. 

In this work we aim to fill this gap by studying large-scale multi-user \ac{mimo} networks utilizing \acp{dma}.  
In particular, we study the achievable performance focusing on the uplink, namely, when data is transmitted from the \acp{ut} to the \ac{bs},  and the \ac{bs} is equipped with a \ac{dma} realizing a large-scale antenna array. 
The application of \acp{dma} results in a simplified hardware, which inherently implements signal processing techniques such as analog combining, subject to specific constraints induced by the physics of the metasurfaces. The resulting structure can be thus used for realizing planar, compact, low cost, and spectral efficient massive \ac{mimo} \acp{bs}. 
Unlike standard analog combining with conventional antenna arrays, e.g., \cite{AlKhateeb:14,Stein:17}, \acp{dma} implement adjustable compression without requiring  additional hardware.

\textcolor{NewColor}{We propose a  model for \ac{mimo} systems with \acp{dma} which encapsulates previously proposed mathematical models for the unique characteristics and constraints of these metasurfaces, such as the frequency response of each metamaterial element \cite{Smith:17,Pendry:99}, the propagation inside the waveguide \cite{Smith:17}, and the mutual coupling induced by the sub-wavelength spacing of the elements \cite{Boyarsky:17,Pulido:16,Pulido:18}. By integrating these established properties of \acp{dma} into the overall \ac{mimo} system model, we obtain an equivalent communication channel including frequency selectivity and constrained linear combining, which can be analyzed using information theoretic tools.}  
Our model also quantifies some of the gains in utilizing \acp{dma}, demonstrating that they require less RF chains compared to standard antenna arrays, thus reducing the cost, memory requirement, and power consumption. 

Next, we focus on the scenario where the wireless channel is frequency flat, and the frequency selectivity, induced by the physics of the metasurfaces, is identical among all the radiating elements. 
We then extend our analysis to the general scenario of frequency selective channels with an arbitrary frequency selectivity profile among the metamaterial elements. For each scenario we characterize the  maximal achievable average sum-rate among all \acp{ut} in the network, and compare it to the fundamental performance limits, which is the maximal achievable sum-rate of frequency selective \ac{mimo} \acp{mac} derived in \cite{Verdu:89}, and requires ideal unconstrained antenna arrays. 
We show that when channel is frequency flat and the frequency selectivity is identical among the elements, its effect can be accounted for in the configuration of the \acp{dma}. Thus, under this setting, when number of \acp{dma} is not smaller than the number of \acp{ut}, \ac{dma} based antenna arrays can approach the fundamental performance limits, achievable using ideal unconstrained antenna arrays. 

For each scenario, we derive an alternating optimization algorithm for configuring the \acp{dma} to approach the performance achievable with unconstrained antenna arrays, accounting for the specific characteristics of the metasurfaces.   
Our numerical analysis demonstrates that the achievable performance of the resulting massive \ac{mimo} systems in which the \ac{bs} implements its large-scale antenna array using \acp{dma} is comparable to the theoretical fundamental limits of the channel. These limits are achievable using unconstrained antenna arrays, which are more costly, require more power and are physically larger compared to \acp{dma} with the same number of radiators.

The rest of this paper is organized as follows: 
Section~\ref{sec:Preliminaries} introduces the mathematical formulation of  \acp{dma}, and defines the problem of uplink multi-user \ac{mimo} communications with \acp{dma}.
Section~\ref{sec:SingleCSI} characterizes the fundamental performance limits achievable with any antenna array, as well as the performance limits when utilizing \ac{dma}, and derives algorithms for designing \acp{dma} to approach the optimal performance.
Section~\ref{sec:Sims} provides simulation examples.
Finally,  Section~\ref{sec:Conclusions}  concludes the paper.
Proofs of the  results are detailed in the appendix. 

Throughout this paper, we use boldface lower-case letters for vectors, e.g., ${\myVec{x}}$;
the $i$th element of ${\myVec{x}}$ is written as $({\myVec{x}})_i$. 
Matrices are denoted with boldface upper-case letters,  e.g., 
$\myMat{M}$,  $(\myMat{M})_{i,j}$   denotes its $(i,j)$th element, ${\rm rank}(\myMat{M})$ denotes its rank and $\left|\myMat{M}\right|$ is its determinant. 
Sets are denoted with calligraphic letters, e.g., $\mathcal{X}$.  
We use $\myI_{n}$ to denote the $n \times n$ identity matrix.  
Stochastic expectation  and mutual information are denoted by  $\E\{ \cdot \}$  and $I\left( \cdot ~ ; \cdot \right)$, respectively. 
%
We use $\left\|\cdot\right\|$ to denote the Euclidean norm when applied to vectors and the Frobenius norm when applied to matrices,
$\otimes$ denotes the Kronecker product,  
   $\mySet{C}$  and $\mySet{N}$ are the sets of   complex numbers  and natural numbers, respectively. 
For any sequence, possibly multivariate, ${\bf y}[i]$,   and integers $b_1 < b_2$,  ${\bf y}_{b_1}^{b_2}$ denotes the column vector obtained by stacking $\Big[\left( {\bf y}[b_1]\right) ^T,\ldots,\left( {\bf y}[b_2]\right) ^T\Big]^T$ and ${\bf y}^{b_2} \equiv {\bf y}_{1}^{b_2}$.

\section{Preliminaries and Problem Formulation}
\label{sec:Preliminaries}
\textcolor{NewColor}{
In the following we model the input-output relationship of \acp{dma} when used on the receive side in a \ac{mimo} communications scenario. This model is based on previously proposed mathematical models for the electromagnetic properties of \acp{dma}, in particular, on the works \cite{Smith:17}  and \cite{Boyarsky:17}. The main contribution of the resulting model lies in its natural integration into the overall communication system model, discussed in the following subsection, allowing the properties of \acp{dma} to be incorporated in an equivalent channel which is analytically tractable from an information theoretic perspective, as shown in Section \ref{sec:SingleCSI}.}
To formulate the considered setup, we first elaborate on metasurface anteannas and mathematically express the input-output relationship of \acp{dma} in Subsection \ref{subsec:Pre_DMA}. Then, we present the massive \ac{mimo} with \acp{dma} system model in Subsection \ref{subsec:Pre_Problem}. Finally, in Subsection \ref{subsec:Pre_Defs} we discuss  the achievable average sum-rate performance metric.

%
\vspace{-0.1cm}
\subsection{Dynamic Metasurface Antennas}
\label{subsec:Pre_DMA}
\begin{figure}
	\centering	
	\includegraphics[trim={0 2cm 0 2cm},width=\figWidth]{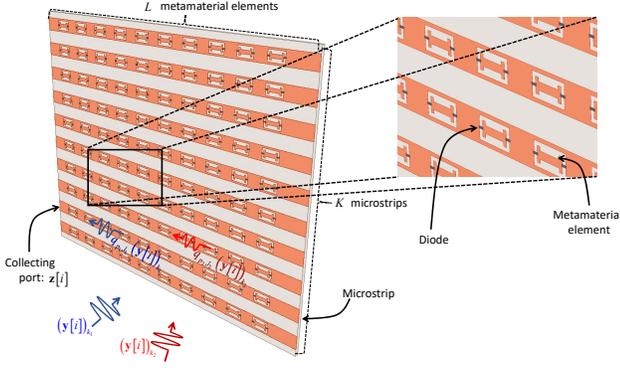}
	\vspace{-0.4cm}
	\caption{Metasurface antenna illustration.}
	\vspace{-0.4cm}
	\label{fig:MetaAntenna}
\end{figure}

Metamaterials are a class of artificial materials whose physical properties, and particularly their permittivity and permeability, can be engineered to exhibit a broad set of desired characteristics \cite{Smith:04,Engheta:06}.  
The underlying idea behind metamaterials is to introduce tailored inclusions in a host medium to emulate a desired effective property. This concept was later extended to surface configurations (thus "metasurface") where the surface effective parameters were tailored to realize a desired transformation on the transmitted, received, or reflected waves \cite{Holloway:12,Pfeiffer:13}. More recently, metasurfaces have been implemented as radiative layers on top of a guiding structure, forming a ``metasurface antenna". In a simple configuration, a metasurface antenna consists of microstrips consisting of a multitude of sub-wavelength, frequency-selective resonant metamaterial radiating elements \cite{Hunt:13}. 
To realize a larger antenna array, such metasurface antennas can be tiled together to form a large array. 
An illustration  of such an array is given in Fig.~\ref{fig:MetaAntenna}.

On the receive side, each microstrip feeds a single RF chain, whose digital output is obtained as a linear combination of the radiation observed by each metamaterial element of the microstrip. This linear combination is a result of the following two physical phenomena: 
\begin{itemize}
	\item \label{txt:FreqSel1} {\em Frequency response of the metamaterial element}: This effect can be typically modeled as a bandpass filter whose quality factor is typically around $30$ \cite{Odabasi:13}, though higher quality factors of around $100$ can also be achieved \cite{Hunt:13}. For example, at carrier frequency of $1.9$ GHz, a quality factor of $30$ would translate into a bandwidth of $63$ MHz. In many relevant communications scenarios, such a response can be considered as frequency flat, namely, the gain induced by the metamaterial element is the same for all the considered frequency range. \textcolor{NewColor}{It is emphasized that this does not imply that the communication channel is frequency flat, as  the  wireless channel gain typically varies in frequency within this band \cite{Kristem:18}.}
	\item {\em Propagation inside the  microstrip}: The effect of this phenomenon depends on the location of the elements along the waveguide (e.g., the microstrip). In particular, by letting $r_{p,l}$ denote the location of the $l$th element on the $p$th microstrip, and $\beta_p$ denote  the wavenumber along the waveguide, which is usually larger than the free space wavenumber $k$, the effect of the propagation inside the waveguide in the frequency domain is proportional to $e^{-j \beta_p \cdot r_{p,l}}$. Since the wavenumber $\beta_p$ is a linear function of the frequency, this effect induces non-negligible frequency selectivity. Thus, we henceforth model this effect in the discrete-time domain as a causal filter with  finite impulse response $\{\Hfilt_{p,l}[\tau]\}_{\tau=0}^{\Mem}$
	\label{txt:taps} \textcolor{NewColor}{whose taps are complex-valued, i.e., $\Hfilt_{p,l}[\tau] \in \mySet{C}$, and $\Mem$ represents the memory of the filter, namely, the number of taps.}
\end{itemize}

It is worth emphasizing here that we have ignored the element-element coupling inside the microstrip for simplicity. This assumption is usually valid when metamaterial elements are weakly coupled to the guided mode \cite{Smith:17}. For cases with strong coupling metamaterial element, one can include such coupling using coupled dipole models \cite{Pulido:16}. This model and its implications (if any) on the massive \ac{mimo} system is beyond the scope of this paper and are left for future works. 

\begin{figure}
	\centering	
	\includegraphics[width=\figWidth]{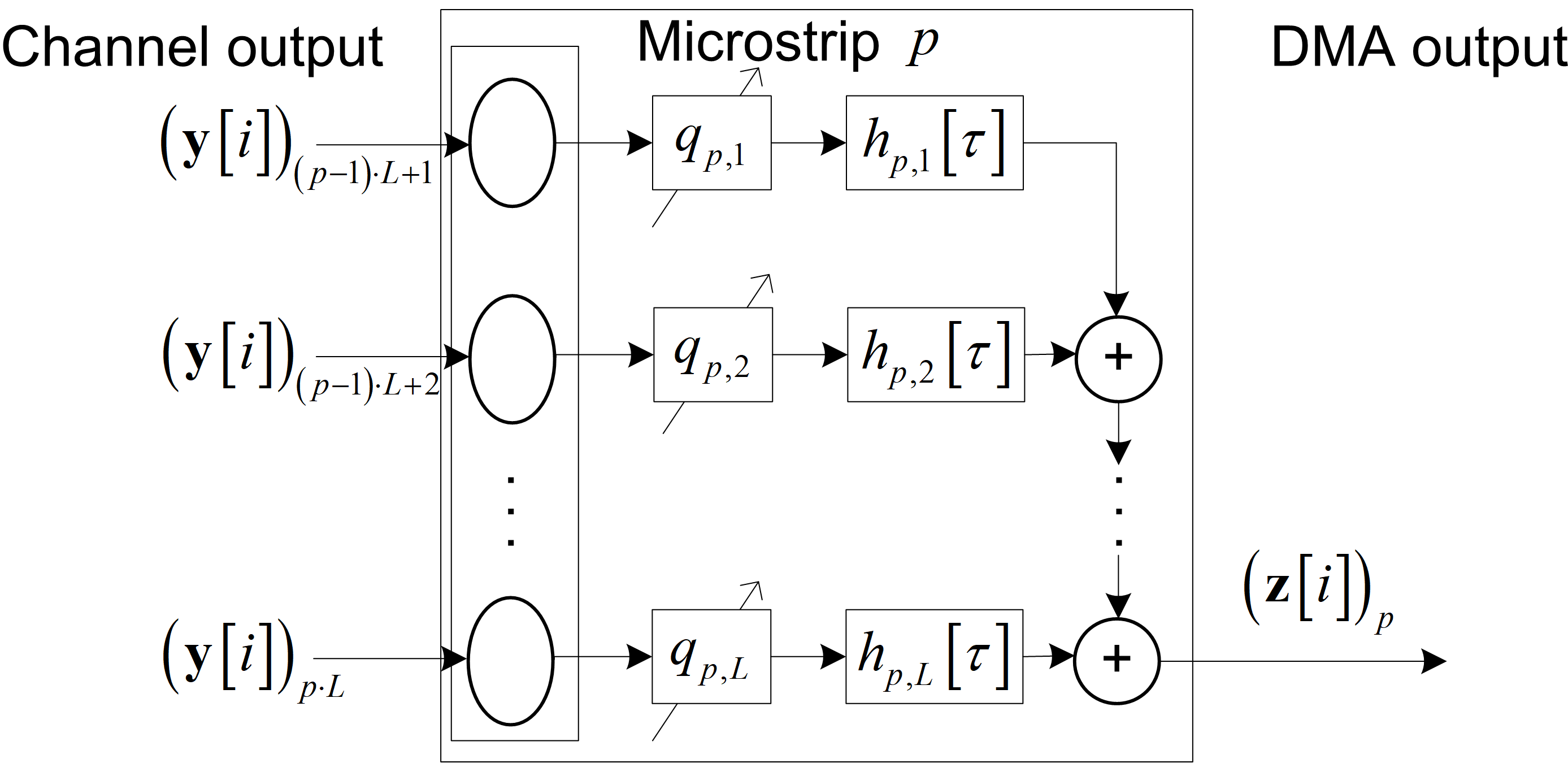}
	\caption{Dynamic metasurface microstrip model.}
	\vspace{-0.4cm}
	\label{fig:DMA_Model1}
\end{figure}

We can now mathematically formulate the input-output relationship of metasurface antennas. Consider a metasurface antenna with $\Ndmas$ microstrips, each consisting of $\Nelements$ elements, and let $\myVec{y}[i] \in \mySet{C}^{\Ndmas\cdot \Nelements \times 1}$ be a vector such that $\left( \myVec{y}[i]\right)_{(p-1) \cdot \Nelements + l}$ denotes the radiation observed at the $l$th element of the $p$th microstrip at time index $i$. 
%
%
The output of the metasurface antenna at time index $i$ is the vector $\myVec{z}[i] \in \mySet{C}^{\Ndmas}$ whose entries can be written as
\begin{equation}
\label{eqn:Metasurface1}
\left( \myVec{z}[i]\right)_p  = \sum\limits_{l=1}^{\Nelements} \myWeightScalar_{p,l}  \sum\limits_{\tau=0}^{\Mem} \Hfilt_{p,l}[\tau]\cdot \left( \myVec{y}[i-\tau]\right) _{(p-1) \cdot \Nelements + l},  
\end{equation}
with $p \in \{1,2,\ldots, \Ndmas\}$. An illustration of the input-output relationship induced by a single manuscript is depicted in Fig. \ref{fig:DMA_Model1}.
It is noted that the filters $\{\Hfilt_{p,l}[\tau] \}$, representing the propagation inside each microstrip, do not depend on the gains of the metamaterial elements $\{ \myWeightScalar_{p,l}\}$, namely, we assume that the metamaterial elements do not perturb the feed wave \cite{Smith:17}.
 Due to the sub-wavelength proximity of the elements in microsrtip, the input vector $\myVec{y}[i]$ is spatially correlated, i.e., its covariance matrix is non-diagonal. We thus do not assume a specific element spacing and incorporate the resulting coupling into the general covariance of $\myVec{y}[i]$.

The relationship between the multivariate processes $\myY[i]$ and  $\myZ[i]$ can  thus be written as 
\begin{equation}
\label{eqn:DMA_Rel1}
\myZ[i] =  \myWeights\sum\limits_{\tau=0}^{\Mem}\Hmat[\tau]\myY[i-\tau], 
\end{equation}
where $\{\Hmat [\tau]\}_{\tau = 0}^{\Mem}$ is a set of $\Nantennas \times \Nantennas$ diagonal matrices, $\Nantennas \triangleq  \Ndmas \cdot \Nelements$, representing the frequency selectivity of the metasurface, i.e., $\left( \Hmat [\tau]\right)_{(p-1)\Nelements +l, (p-1)\Nelements +l} = \Hfilt_{p,l}[\tau]$, and $\myWeights$ is an $\Ndmas \times \Nantennas $ matrix representing the configurable weights of the \acp{dma}. Using  \eqref{eqn:Metasurface1}, we can write 
\begin{equation}
\label{eqn:Qmat1}
\left( \myWeights\right)_{p_1, (p_2-1)\Ndmas +l  } = 
\begin{cases}
\myWeightScalar_{p_1, l} & p_1 = p_2 \\
0 & p_1 \ne p_2
\end{cases} .
\end{equation}
{For mathematical convenience, it is assumed that the coefficients $\{\myWeightScalar_{p,l}\}$ are unitless, i.e.,  the polarizability of the elements is normalized.}
\textcolor{NewColor}{
While the model detailed above considers a \ac{dma} in which the radiating elements are placed along a set of one-dimensional microstrip, it can incorporate a broader family of two-dimensional \acp{dma}. In fact, any two-dimensional \ac{dma} in which each element is connected to a single output port can be represented via \eqref{eqn:DMA_Rel1} by modifying the structure of the weights matrix $\myWeights$ to represent the resulting elements interconnections.}

\acp{dma} integrate a tuning mechanism into each independent resonator of a metasurface antenna \cite{Sleasman:16}. The dynamic tuning adds the flexibility to adjust the properties of the metamaterial elements, namely, to control the values of the coefficients  $\{\myWeightScalar_{p,l}\}$  in \eqref{eqn:Metasurface1}. The set of possible values of $\{\myWeightScalar_{p,l}\}$, denoted $\myWeightSet$, represents the Lorentzian resonance response \cite{Smith:17}, and typically consists of
a subset of the complex plane $\mySet{C}$ of either of the following forms \cite[Sec. III]{Smith:17}: 
\begin{itemize}
	\item {\em Amplitude only}, namely, $\myWeightSet = [a,b]$ for some real non-negative $a < b$.
	\item {\em Binary amplitude}, i.e., $\myWeightSet = c \cdot \{0,1\}$ for some fixed $c \in \mySet{R}^+$.
	\item {\em Lorentzian-constrained phase}, that is $\myWeightSet = \{q=\frac{j + e^{j \phi}}{2}: \phi \in [0, 2\pi]  \}$.
\end{itemize}

\label{txt:Gains}
\textcolor{NewColor}{
In order to quantify the gains of utilizing \acp{dma}, we next compare these antenna architectures to standard antenna arrays. We use the term {\em standard arrays} for systems where the receiver is capable of directly processing the observed vector $\myY[i]$, which is the common model in the \ac{mimo} communications literature, for both conventional \ac{mimo} \cite[Ch. 7]{Tse:05} as well as massive \ac{mimo}  \cite{Marzetta:10, Hoydis:13,Shlezinger:17}.
	Clearly, any performance achievable with \ac{dma}-based antenna arrays is also achievable with standard antenna arrays, as $\myZ[i]$ can be obtained from $\myY[i]$, but not vice versa. 
	However, unless an additional RF chain reduction hardware is used, such as analog combiners discussed in the sequel, standard antenna arrays require each of the $\Ndmas \cdot \Nelements$ radiating elements to be connected to an RF chain as well as an \ac{adc}, while \acp{dma} require a single RF chain and \ac{adc} per microstrip.}  Note that RF chain hardware tends to be costly \cite{Rial:16}, and that \acp{adc} are typically a dominant source of power consumption \cite{Mo:17} and memory usage \cite{Shlezinger:18}. Consequently, by utilizing \acp{dma}, the resulting cost, memory usage, and power consumption, are reduced by a factor of $\Nelements$ compared to standard antenna arrays \cite{Yoo:18}. 
\label{txt:Geometry}
\textcolor{NewColor}{Additionally, \acp{dma}  can realize planar antenna arrays \cite{Mikala:15,Mikala:16}, and unlike standard antenna arrays, they can pack a larger number of elements into a given physical area \cite{Akyildiz:16}.}

\textcolor{NewColor}{We note that reducing the number of RF chains and \acp{adc} can also be carried out with standard antenna arrays using dedicated analog combining hardware, see, e.g., \cite{Rial:16,AlKhateeb:14,Stein:17,Shlezinger:18}. However, in the presence of standard antenna arrays, analog combining comes at the cost of additional hardware, which increases the overall size and cost, especially when the analog combining should be adjustable in run-time. The exact value of the increased cost and size depends on the specific implementation of the analog combiner. \acp{dma} inherently implement adjustable analog combining in the physical structure of the metasurfaces without additional hardware. Specifically, multiple metamaterial radiators are fed directly by a waveguide structure to simplify the feed structure, avoiding the usage of potentially more expensive and complicated RF circuitry. Thus, \acp{dma} have been recognized as a radiative platform with a simple, energy-efficient, low-cost, and low-profile configuration.}
Furthermore, standard analog combining implemented using dedicated hardware is typically subject to different constraints than those imposed on $\myWeights$ here. In particular, while in \acp{dma} the weights matrix $\myWeights$ must obey the structure in \eqref{eqn:Qmat1} and its entries must be in the feasible set $\myWeightSet$ defined above, standard analog combiners must satisfy the architecture-based constraints detailed in \cite[Sec. II]{Rial:16}, such as the commonly used {\em phase shifting network} constraint, i.e., $\myWeightSet = \{q\in \mySet{C}: |q| = 1  \}$, or the {\em switching network} constraint, in which  $\myWeightSet =  \{0,1\}$.
It is also noted that when $\Nelements = 1$, $\myWeights = \myI_{\Ndmas}$, and $\{\Hfilt_{p,l}[\tau]\}$ are Kronecker delta functions, namely, each microstrip realizes a single frequency flat antenna, then $\myZ[i] \equiv \myY[i]$, and the resulting \ac{dma} coincides with the standard antenna array. However, this implementation requires the same amount of RF chains and \acp{adc} as standard arrays, and thus does not result in any gains in terms of cost, power consumption, and memory requirement.
\textcolor{NewColor}{Finally, it is emphasized that while we consider \acp{dma} with frequency flat element response, resulting in the frequency invariant weights matrix $\myWeights$, it is possible to design \acp{dma} to have dynamically adjustable frequency selective weights. This can be achieved by using elements with different resonance frequency along the microstrip and turning them on and off to realize a desired frequency selective response. While designing frequency selective \acp{dma} is expected to introduce an additional potential gain over conventional analog combiners, 	the set of possible frequency selectivity profiles which can be realized in \acp{dma}
	is heavily implementation dependent, and as a result, we leave this for future investigation.}

\label{txt:DMASum}
\textcolor{NewColor}{
	To summarize, \acp{dma} realize antenna arrays with specific structure constraints, representing the underlying physics of the metasurface. These constraints include additional filtering of the received signal due to the propagation inside the microstrip; spatial correlation due to sub-wavelength element spacing; and an inherent adjustable signal compression as the signals are combined in each microstrip. The benefits of using \acp{dma} as an antenna array architecture are their low-cost, power-efficiency, and physical shape and size flexibility. An additional benefit in the context of massive \ac{mimo} communications, which we exploit in the sequel, is their natural ability to implement a form of dynamic analog combining as an integral part of the antenna structure without requiring additional dedicated hardware.
	}

\subsection{System Model}
\label{subsec:Pre_Problem}
We consider a noncooperative single-cell multi-user \ac{mimo} system, focusing on the uplink. 
The \ac{bs} is equipped with a \ac{dma}, consisting of $\Ndmas$ microstrips, each with $\Nelements$  elements, namely, the overall number of radiating elements used by the \ac{bs} is $\Nantennas \triangleq  \Ndmas \cdot \Nelements$. The number of \acp{ut} served by the \ac{bs} is $\Nusers$, assumed to be not larger than $\Nantennas$.

Let  $\{\Gmat[\tau]\}_{\tau=0}^{\MemG}$ be a set of ${\Nantennas \times \Nusers}$ matrices representing the multipath channel  matrix from the \acp{ut}  to the  \ac{bs}, where $\MemG$ denotes the length memory of the discrete-time channel transfer function, i.e., the number of taps is $\MemG +1$, and $\MemG = 0$ implies that the channel is memoryless. 
The channel output at the \ac{bs} is corrupted by  be an i.i.d. zero-mean proper-complex multivariate Gaussian noise  $\myW[i]\in \mySet{C}^{\Nantennas}$ with  covariance matrix $\CovW$. 
By letting $\myX[i]\in\mySet{C}^{\Nusers}$ be the transmitted signal of the \acp{ut} at time index $i$, the corresponding channel output at the \ac{bs} is given by
\begin{equation}
\label{eqn:Channel_Rel1}
\myY[i] = \sum\limits_{\tau = 0}^{\MemG}\Gmat[\tau]\myX[i - \tau] + \myW[i].
\end{equation}  
We assume that the \acp{ut} utilize Gaussian codebooks, i.e., $\myX[i]$ is a zero-mean Gaussian vector with identity covariance matrix, and that the \ac{bs} has full \ac{csi}, namely, the matrices $\{\Gmat[\tau]\}_{\tau=0}^{\MemG}$ are known to the \ac{bs}.

\begin{figure}
	\centering
	\includegraphics[width=\figWidth]{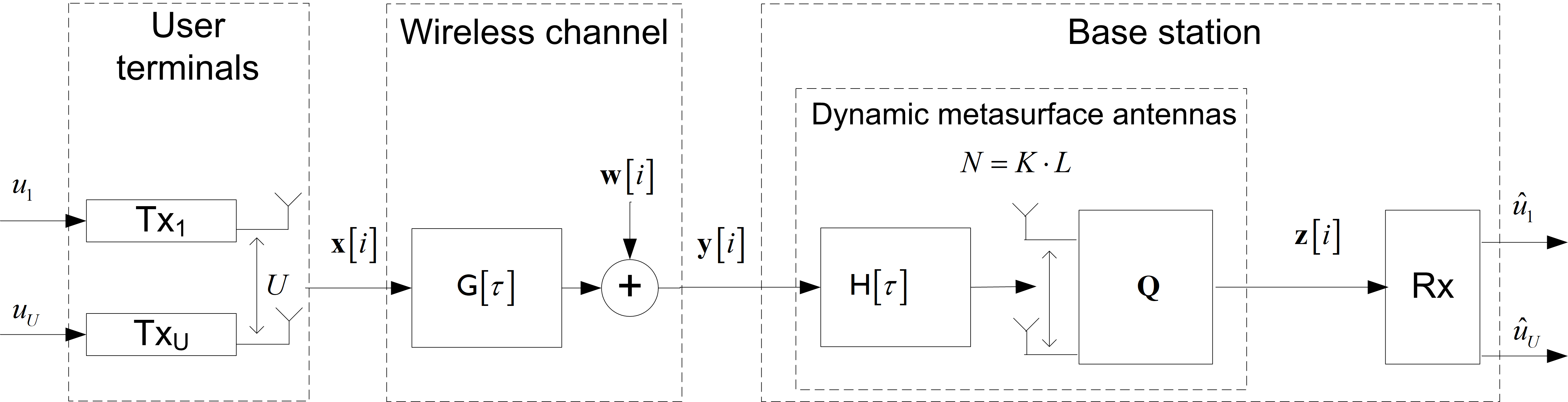}
	\caption{System model illustration.}
	\vspace{-0.4cm}
	\label{fig:SingleCell}
\end{figure}

At the \ac{bs}, the \ac{dma} converts the received signal $\myY[i] \in \mySet{C}^{\Nantennas}$ into the vector $\myZ[i] \in \mySet{C}^{\Ndmas}$, which is used to decode the transmitted signal. As detailed in Subsection \ref{subsec:Pre_DMA},  the relationship between $\myY[i]$ and  $\myZ[i]$ is given by \eqref{eqn:DMA_Rel1}. 
Note that the frequency selectivity of the metasurface is modeled via $\{\Hmat [\tau]\}_{\tau = 0}^{\Mem}$. An illustration  is given in Fig. \ref{fig:SingleCell}.

We note that unlike the standard massive \ac{mimo} literature, e.g., \cite{Marzetta:10,Hoydis:13}, which models the channel as memoryless, the model in \eqref{eqn:Channel_Rel1} explicitly accounts for the frequency selectivity of the wireless channel.  
Frequency selective \ac{mimo} models as in \eqref{eqn:Channel_Rel1}, i.e., without the presence of \acp{dma}, were studied for point-to-point communications \cite{Brandenburg:74}, \acp{mac} \cite{Verdu:89}, broadcast channels \cite{Goldsmith:01}, and wiretap channels \cite{Shlezinger:16}.
 Our motivation for using the model in \eqref{eqn:Channel_Rel1} stems from the fact that the standard massive \ac{mimo} memoryless model is obtained by assuming   \ac{ofdm} modulation with subcarrier spacing smaller than the coherence bandwidth of the channel, and cyclic prefix of length larger than the number of multipath taps \cite[Ch. 3]{Tse:05}. However, when metasurfaces are present, it is no longer reasonable to assume that the  moderate frequency variations induced by the \ac{dma} are effectively canceled by the transmission scheme.  Consequently, in order to account for the frequency selectivity of both the \ac{dma} as well as the wireless channel, we explicitly incorporate the effect of multipath into the model in~\eqref{eqn:Channel_Rel1}.  
 


In order to compare the performance achievable with \ac{dma} to that achievable with ideal unconstrained antenna arrays, we also consider the case where the \ac{bs} decodes the transmitted signals based on the output of the wireless channel $\myY[i]$, instead of $\myZ[i]$. This scenario is referred to henceforth as {\em optimal \ac{mimo}}.   As discussed in Subsection \ref{subsec:Pre_DMA}, the maximal \ac{se} of {optimal \ac{mimo}} is not smaller\footnote{To be precise, as the considered channel is {\em information stable}, the achievable sum-rate can be expressed using the mutual information between its input and output \cite[Ch. 4]{ElGamal:11}. Therefore, as $\myZ[i]$ is a deterministic mapping of $\myY[i]$, it follows from the data processing inequality \cite[Ch. 2]{ElGamal:11} that the achievable sum-rate when the output  $\myY[i]$ is not smaller than that achievable when the output is $\myZ[i]$. } than that achievable with \acp{dma}, as the output of the \ac{dma} $\myZ[i]$ can be obtained from $\myY[i]$. 
Since different configurations of the same number of elements results in a different statistical model for the channel output, in order to maintain fair comparison, the antenna spacing between the $\Nantennas$ elements used in the optimal \ac{mimo} setup is identical to that used with \acp{dma}. Under this setting, the resulting wireless channel, namely, the relationship between $\myX[i]$ and $\myY[i]$, is the same as in the \ac{dma} setup.

Our goal is to characterize the performance achievable for the considered system with \acp{dma} compared to the optimal \ac{mimo} case, and to provide guidelines for configuring the \ac{dma} weights such that the performance is optimized. In the following section we properly define the performance metric used henceforth.

\vspace{-0.2cm}
\subsection{Definitions}
\label{subsec:Pre_Defs}
\vspace{-0.1cm}
In order to rigorously formulate the performance metric used in the paper, while accounting for the frequency selectivity induced by the \acp{dma}, we present a set of necessary definitions, which are based on \cite[Ch. 4]{ElGamal:11}. We begin with the definition of finite-memory multi-user channels:
\vspace{-0.1cm}
\begin{definition}[A multi-user \ac{mimo} channel with finite-memory]
	\label{def:Channel}
	A discrete-time $\Nantennas \times \Nusers$ multi-user \ac{mimo} channel with finite memory
	consists of a set of $\Nusers$ scalar input sequences, represented via a multivariate sequence $\myX[i] \in \mySet{R}^{\Nusers}$, $i \in \mySet{N}$, an output sequence  $\myY[i] \in \mySet{R}^{\Nantennas}$, $i \in \mySet{N}$, an initial state vector $\myVec{s}_0 \in \mySet{S}_0$ of finite dimensions, and a sequence of conditional probabilities $\big\{p\left(\myY^{n}|\myX^{n},\myVec{s}_0\right)\big\}_{n=0}^{\infty}$. 
\end{definition}

Having defined a multi-user \ac{mimo} channel with finite-memory, we can now introduce the definition of codes for such channels:
\begin{definition}[Multi-user code]
	\label{def:Code}
	A $\left[\{R_k\}_{k=1}^{\Nusers}, l \right]$ multi-user code with rates $\{R_k\}_{k=1}^{\Nusers}$ and blocklength $l \in \mySet{N}$ consists of:
	
	{\em 1)} $\Nusers$ message sets $\mySet{U}_k \triangleq \{1,2,\ldots,2^{lR_k}\}$, with $k \in \{1,2,\ldots, \Nusers\}$.
	
	{\em 2)} A  family of encoders $e_{l,k}$, each maps the message $u_k \in \mySet{U}_k$ into a codeword $\myX_{(u_k)}^{l} = \big[x_{(u_k)}\left[1\right],\ldots,x_{(u_k)}\left[l\right]\big]^T$, i.e.,
	\begin{equation*}
	e_{l,k}:\mySet{U}_k \mapsto \mySet{R}^l.
	\end{equation*}
	The channel input is $\myX[i] = \big[x_{(u_1)}\left[i\right], \ldots,x_{(u_{\Nusers})}\left[i\right] \big]^T$.
	
	{\em 3)}  A decoder $d_l$ which maps the channel output  $\myY^{l}$ into the messages $\hat{u}_1,\ldots \hat{u}_{\Nusers} $, i.e.,
	\begin{equation*}
	d_l: \mySet{R}^{\Nantennas \times l} \mapsto  \mySet{U}_1 \times \cdots \times \mySet{U}_{\Nusers}.
	\end{equation*}
	The encoders and decoder operate independently of the initial state $\myVec{s}_0$.
	
\end{definition}
The set $\big\{x_{(u_k)}^{l}\big\}_{u =1}^{2^{lR_k}}$ is referred to as the $k$-th {\em codebook} of the $\left[\{R_k\}_{k=1}^{\Nusers}, l \right]$ code.
Assuming each message $u_k$ is uniformly selected from $\mySet{U}_k$, the average probability of error, when the initial state is $\myVec{s}'_0$, is given by \cite[Sec. III]{Goldsmith:01}:
\ifsingle
\begin{equation}
\label{eqn:def_avgError}
P_{\rm e}^l \left( \myVec{s}'_0 \right) \!= \! \frac{1}{2^{l \sum\limits_{k=1}^{\Nusers} R_k}}\sum\limits_{u_1 \!= \! 1}^{2^{lR_1}} \cdots \sum\limits_{u_{\Nusers} \!= \! 1}^{2^{lR_{\Nusers}}} \Pr \left( {\left. {{d_l }\left( \myY^{l}  \right) \ne \big[u'_1, \ldots, u'_{\Nusers}\big] } \right|u_1\!=\!u'_1, \ldots, u_{\Nusers}\!=\! u'_{\Nusers},{\myVec{s}_0} \!= \! {\myVec{s}'_0}} \right).
\end{equation}
\else
\begin{align}
P_{\rm e}^l \left( \myVec{s}'_0 \right) \!= \! \frac{1}{2^{l \sum\limits_{k=1}^{\Nusers}\! R_k}}\sum\limits_{u_1 \!= \! 1}^{2^{lR_1}}\! \cdots\! \sum\limits_{u_{\Nusers} \!= \! 1}^{2^{lR_{\Nusers}}} \Pr \Big(&  {d_l }\left( \myY^{l}  \right) \ne \big\{u'_i \big\}   \notag \\
&\Big|\big\{u_i\!=\!u'_i\big\} ,{\myVec{s}_0} \!= \! {\myVec{s}'_0} \Big).
\label{eqn:def_avgError}
\end{align}
\fi 

Using Definition \ref{def:Code}, we can now properly formulate the definition for achievable sum-rate, which will be used henceforth as our main metric for evaluating multi-user \ac{mimo} networks operating with \acp{dma}. 
\begin{definition}[Achievable average sum-rate]
	\label{def:Rate}
	An average sum-rate $\Rs$ is called achievable if, for every $\epsilon_1,\epsilon_2 > 0$, there exists a positive integer $l_0 >0$ such that for all integer $l > l_0$, there exists a multi-user code, $\left[\{R_k\}_{k=1}^{\Nusers}, l \right]$,  which satisfies 
	\begin{subequations}
		\label{eqn:DefRate}
		\begin{equation}
		\label{eqn:DefRate1}
		\mathop {\sup }\limits_{{\myVec{s}'_0} \in \mathcal{S}_0} P_{\rm e}^l \left( {{\myVec{s}'_0}} \right) < {\epsilon _1},
		\end{equation}
		and
		\begin{equation}
		\label{eqn:DefRate2}
		\frac{1}{\Nusers}\sum\limits_{k=1}^{\Nusers} R_k \geq \Rs - \epsilon_2.
		\end{equation}
	\end{subequations}
\end{definition} 

Note that the achievable average sum-rate in Definition \ref{def:Rate} is a {\em fundamental property of the multi-user \ac{mimo} channel}. In fact, for a given set of multi-user encoders, the decoder which maximizes the achievable average sum-rate typically implements joint decoding \cite[Ch. 4]{ElGamal:11}. Such decoders are usually computationally complex, and thus many works on massive \ac{mimo} focus on the achievable average sum-rate assuming less complex suboptimal separate linear decoding, see  \cite{Marzetta:10,Hoydis:13,Shlezinger:17}. 
%
\textcolor{NewColor}{In this work we focus on the implementation of the antenna array using \acp{dma}, with the promise of reducing cost, size, and power consumption. Consequently, we impose no constraints on the processing and decoding carried out in the {\em digital domain}, assume that the \ac{bs} has perfect knowledge of the underlying channel, and characterize the performance in terms of the maximal achievable average sum-rate. It is emphasized that operating under computational complexity constraints and obtaining an accurate channel estimation are challenging tasks on their own for massive \ac{mimo} \acp{bs} with \acp{dma}. We consider the design of efficient decoding and channel estimation schemes, as well the analysis of the effect of inaccurate channel knowledge and hardware impairments on the resulting performance when using \acp{dma}, as potential future research directions, extending the current study.}

\section{Achievable Average Sum-Rates}
\label{sec:SingleCSI}
In the following we study the achievable average sum-rate and the resulting \ac{dma} configuration for the setup presented in Subsection \ref{subsec:Pre_Problem}. 
%
To formulate the achievable average sum-rate, let  $\HmatFD(\omega)$ and $\GmatFD(\omega)$, $\omega \in [0,2\pi)$, denote the  \acp{dtft} of ${\Hmat}[\tau]$ and of $\Gmat[\tau]$, respectively. The maximal achievable average sum-rate for a fixed \ac{dma} weights matrix $\myWeights$ is stated in the following theorem:

\begin{theorem}
	\label{thm:SumRate1}
	The maximal achievable average sum-rate  of the channel in \eqref{eqn:Channel_Rel1} and \eqref{eqn:DMA_Rel1} for a fixed weight matrix $\myWeights$ is given by
	\ifsingle
	\begin{align}
	\Rs \! =\! \frac{1}{2 \pi \cdot \Nusers}\mathop{\int}\limits_{0}^{2\pi} \log \bigg| \myI_{\Ndmas} +  \myWeights {\HmatFD}(\omega) \GmatFD(\omega)\GmatFD^H(\omega) {\HmatFD}^H(\omega)\myWeights^H   \left(  \myWeights {\HmatFD}(\omega) \CovW{\HmatFD}^H(\omega)\myWeights^H  \right)^{-1}\!  \bigg| d \omega.
	\label{eqn:SumRate1}
	\end{align}	
	\else
	\begin{align}
	\Rs \! =\! \frac{1}{2 \pi \cdot \Nusers}\mathop{\int}\limits_{0}^{2\pi} \log \bigg|&\myI_{\Ndmas} +  \myWeights {\HmatFD}(\omega)  \GmatFD(\omega)\GmatFD^H(\omega) {\HmatFD}^H(\omega)\myWeights^H \notag \\
	&\times \left(  \myWeights {\HmatFD}(\omega) \CovW{\HmatFD}^H(\omega)\myWeights^H  \right)^{-1}\!  \bigg| d \omega.
	\label{eqn:SumRate1}
	\end{align}
	\fi 
\end{theorem}

{\em Proof:}
See Appendix \ref{app:Proof1}.

\smallskip
Theorem \ref{thm:SumRate1} characterizes the maximal achievable sum-rate by incorporating the \ac{dma} operation as part of the channel, and obtaining the achievable sum-rate of the resulting finite-memory \ac{mac} as in \cite{Verdu:89}. 
Theorem \ref{thm:SumRate1} can also be used to obtain the fundamental performance limits of the wireless channel, achievable with optimal unconstrained antenna arrays, as stated in the following corollary:
\begin{corollary}
	\label{cor:OptMIMO}
	 Define   $\EqGMatFD(\omega) \triangleq \CovW^{-1/2}  \GmatFD(\omega)\GmatFD^H(\omega)\CovW^{-1/2}$,  and let $\{\lambda_i(\omega)\}_{i=1}^{\Nusers}$ be its eigenvalues arranged in descending order. The maximal achievable average sum-rate of the optimal \ac{mimo} setup is given by
	 	\begin{equation}
	 	\label{eqn:OptMIMO}
	 	\Rs\Mopt = \frac{1}{2 \pi \cdot \Nusers}\mathop{\int}\limits_{0}^{2\pi}\sum\limits_{i=1}^{\Nusers}\log\big(1+\lambda_i(\omega)\big) d \omega.
	 	\end{equation}
\end{corollary}

\begin{IEEEproof}
	As noted in Subsection \ref{subsec:Pre_DMA}, when $\Nelements = 1$, $\myWeights = \myI_{\Ndmas}$, and ${\HmatFD}(\omega) \equiv \myI_{\Ndmas}$, the resulting setup coincides with the optimal \ac{mimo} setup. Substituting this into \eqref{eqn:SumRate1} proves \eqref{eqn:OptMIMO}. 
\end{IEEEproof}

\smallskip
When \acp{dma} are utilized, we note that due to the integration operation and the structure constraints on $\myWeights$, it is difficult to determine the \ac{dma} weights matrix $\myWeights$ such that \eqref{eqn:SumRate1} is maximized.  Therefore, in order to design $\myWeights$ and obtain the resulting $\Rs$, we first   focus on the special case where all the metasurface elements exhibit the same frequency selectivity profile, and the wireless channel is frequency flat. For this case, we derive in Subsection \ref{subsec:OptimalIFreq} the choice of $\myWeights$ which maximizes the achievable sum-rate, ignoring the structure constraints detailed in Subsection \ref{subsec:Pre_DMA}. Then, in Subsection \ref{subsec:PracticalIFreq} we propose an iterative algorithm for configuring practical constrained \acp{dma}. Finally, in Subsection \ref{subsec:PracticalAFreq}, we show how these design principles can be extended to arbitrary frequency selectivity profiles.

\subsection{Optimal Weights for Flat Channels with Identical Frequency Selectivity}
\label{subsec:OptimalIFreq}
The maximal achievable average sum-rate and the corresponding weights configuration $\myWeights$ which maximizes \eqref{eqn:SumRate1} are in general difficult to compute. Thus, we will first consider the special case  where all the metamaterial elements exhibit the same frequency selectivity profile, and the wireless channel is frequency flat. Under this model, the multivariate filter representing the response of the antennas $\Hmat[\tau]$ can be written as 
$\Hmat[\tau]=\myI_{n_t}\cdot h[\tau]$, for some scalar mapping $h[\tau]$, and the multipath channel is given by a single tap $\Gmat = \Gmat[0]$, i.e., $\MemG = 0$. 
By letting $\gamma(\omega)$ be the \ac{dtft} of $h[\tau]$, the multivariate \acp{dtft} of  $\Hmat[\tau]$ and of $\Gmat[\tau]$  can be written ${\HmatFD}(\omega) = \myI_{n_t}\cdot\gamma(\omega)$  and $\GmatFD(\omega) = \Gmat$. Thus,  the achievable average sum-rate in \eqref{eqn:SumRate1} is given by
\begin{equation}
\label{eqn:Case1SE}
\Rs =  \frac{1}{\Nusers}  \log \left|\myI_{\Ndmas} +
\myWeights  \Gmat\Gmat^H
  \myWeights^H \left(  \myWeights  
\CovW  \myWeights^H  \right)^{-1}  \right|.
\end{equation}

In order to find  $\myWeights$ which maximizes \eqref{eqn:Case1SE}, we define $\EqGMat \triangleq \CovW^{-1/2}  \Gmat\Gmat^H\CovW^{-1/2}$. We now  formulate the dependence of $\Rs$ on $\myWeights$ in the following lemma:
\begin{lemma}
	\label{lem:RelQRs}
	Define $\tilde{\myWeights} \triangleq \myWeights \CovW^{1/2}$ and let $\myMat{V}$ be its right singular vectors matrix. By letting $\tilde{\myMat{V}}$ be the $\Nantennas\times \Ndmas$ matrix consisting of the first $\Ndmas$ columns of the unitary matrix $\myMat{V}$, the achievable sum-rate in \eqref{eqn:Case1SE} can be written as
	\begin{equation}
	\label{eqn:RelQRs}
	\Rs = \frac{1}{\Nusers}  \log \left|\myI_{\Ndmas} +
	\tilde{\myMat{V}}^H \EqGMat \tilde{\myMat{V}}  \right|.
	\end{equation}
\end{lemma}  

\begin{IEEEproof}
	By replacing $\myWeights$ in \eqref{eqn:Case1SE} with $\tilde{\myWeights} = \myWeights \CovW^{1/2}$ it follows from Sylvester's determinant theorem \cite[Ch. 6.2]{Meyer:00} that 
	\begin{equation}
	\Rs =  \frac{1}{\Nusers} \log \left|\myI_{\Nantennas} + \EqGMat\left( \tilde{\myWeights}^H \left( \tilde{\myWeights} \tilde{\myWeights}^H \right)^{-1}\tilde{\myWeights} \right) 
	\right|.
	\label{eqn:ProofLem1A}
	\end{equation}
	Next, we note that $\tilde{\myWeights}^H \left( \tilde{\myWeights} \tilde{\myWeights}^H \right)^{-1}\tilde{\myWeights} $ is a projection matrix, and can therefore be written as $\tilde{\myWeights}^H \left( \tilde{\myWeights} \tilde{\myWeights}^H \right)^{-1}\tilde{\myWeights}  = \tilde{\myMat{V}} \tilde{\myMat{V}}^H$ \cite[Ch. 5.9]{Meyer:00}. Substituting this into \eqref{eqn:ProofLem1A}  proves \eqref{eqn:RelQRs}.
\end{IEEEproof}

Lemma \ref{lem:RelQRs} implies that the achievable average sum-rate $\Rs$ depends on the weights matrix $\myWeights$ {\em only through the first $\Ndmas$ right eigenvectors} of $\tilde{\myWeights} = \myWeights \CovW^{1/2}$. If we ignore the structure constraints on $\myWeights$, then the maximal achievable sum-rate and the corresponding choice of $\tilde{\myMat{V}}$ which maximizes \eqref{eqn:ProofLem1A} are given in the following corollary:
\begin{corollary}
	\label{cor:OptSetting}
 Let $\{\lambda_i\}_{i=1}^{\Nusers}$ be the eigenvalues of $\EqGMat$  arranged in descending order.
	Then, the maximal achievable average sum-rate when $\myWeights$ can be any complex matrix is given by
	\begin{equation}
	\label{eqn:OptSetting}
	\Rs\opt = \frac{1}{\Nusers}\sum\limits_{i=1}^{\min\left( \Ndmas, \Nusers\right) }\log(1+\lambda_i).
	\end{equation}
	The rate $\Rs\opt$ achieved by setting the columns of $\tilde{\myMat{V}}$ to be the eigenvectors  corresponding to  $\{\lambda_i\}_{i=1}^{\Nusers}$.
\end{corollary}

\begin{IEEEproof}
	The corollary follows directly from \eqref{eqn:ProofLem1A}. 
\end{IEEEproof}

\smallskip
Note that the number of non-zero eigenvalues of $\EqGMat$ is equal to its rank, denoted $\NGmat$, which  is at most $\Nusers$. It therefore follows from \eqref{eqn:OptSetting} that increasing the number of microstrips $\Ndmas$ to be larger than $\NGmat$ has no effect on the optimal sum-rate $\Rs\opt$. 
In particular, comparing \eqref{eqn:OptSetting} to the fundamental limits in \eqref{eqn:OptMIMO}, we note that when $\Ndmas \ge \NGmat$, then $\Rs\opt$ achieves the fundamental limits $\Rs\Mopt$. 
However, as each microstrip requires a single RF chain and \ac{adc}, increasing $\Ndmas$ implicitly increases the cost, power usage, and memory requirements of the resulting system.  
Furthermore, by letting $\tilde{\myMat{U}} \tilde{\myMat{D}} \tilde{\myMat{V}}^H$ be the compact \ac{svd} of the optimal $\tilde{\myWeights}$, it follows from Corollary \ref{cor:OptSetting} that the weights matrix which maximizes \eqref{eqn:Case1SE} can be written as
\begin{equation}
\label{eqn:OptimalQ}
\myWeights\opt = \tilde{\myMat{U}} \tilde{\myMat{D}}\tilde{\myMat{V}}^H \CovW^{-1/2}.
\end{equation} 
In particular, the matrix in \eqref{eqn:OptimalQ} maximizes \eqref{eqn:Case1SE} for any setting of unitary $\Ndmas \times \Ndmas$ matrix $\tilde{\myMat{U}}$ and diagonal  $\Ndmas \times \Ndmas$ matrix $\tilde{\myMat{D}}$ with positive diagonal entries. 

It follows from \eqref{eqn:OptimalQ} that the optimal weights matrix $\myWeights\opt$ implements the following processing: First, it applies a noise whitening filter, modeled via the matrix $\CovW^{-1/2}$. 
Then, it utilizes the transformation $\tilde{\myMat{V}}^H$ to project the output into its least noisy $\Ndmas \times 1$ subspace, determined by the largest singular values of the whitened channel transfer matrix $\CovW^{-1/2}\Gmat$, or alternatively, by the largest eigenvalues of $\EqGMat$.  
The fact that  $\myWeights\opt$ depends on the channel and the statistics of the noise indicates that the ability to reconfigure the analog combining weights, which is inherently supported by \acp{dma}, is vital in wireless communications. 
Finally, we note that the remaining invertible processing, determined by the matrices $\tilde{\myMat{U}}, \tilde{\myMat{D}}$, has no effect on the resulting achievable rate, in agreement with the data processing inequality \cite[Ch. 2.3]{ElGamal:11}. However, in the following subsection we show that these matrices can be used to facilitate the approximation of  $\myWeights\opt$ via a feasible weights matrix, which satisfies \eqref{eqn:Qmat1} and whose entries belong to $\myWeightSet$.

\subsection{Practical Design for Flat Channels with Identical Frequency Selectivity}
\label{subsec:PracticalIFreq}
The derivation of the optimal sum-rate in Corollary \ref{cor:OptSetting} ignores the structure constraints on $\myWeights$, and assumes that the right eigenvectors matrix $\tilde{\myMat{V}}$ can be any set of unitary vectors. Nonetheless, as detailed in the problem formulation in Subsection \ref{subsec:Pre_Problem}, $\myWeights$  must be written as in \eqref{eqn:Qmat1}, and  its coefficients $\{\myWeightScalar_{i,l}\}$ should belong to the feasible set $\myWeightSet$. Since finding the constrained matrix $\myWeights$ which maximizes \eqref{eqn:Case1SE} is a difficult task, we propose to set $\myWeights$ to be the closest feasible matrix to the unconstrained  $\myWeights\opt$ in the sense of minimal Frobenious norm. Here, as in \cite{Stein:17, Shlezinger:17b}, we exploit the fact that $\Rs\opt$ is invariant to the selection of the left singular matrix $\tilde{\myMat{U}}$ and the diagonal singular values matrix $\tilde{\myMat{D}}$, and set these matrices such that the Frobenious distance to the feasible approximation is minimized. 
To formulate the problem, we let $\myWeightSet^{\Ndmas \times \Nantennas}$ be the set of $\Ndmas \times \Nantennas$ which can be written as in \eqref{eqn:Qmat1} and whose non-zero entries belong to the feasible set $\myWeightSet$. Let $\mySet{U}^{\Ndmas}$ denote the set of  $\Ndmas \times \Ndmas$ unitary matrices, and $\mySet{D}^{\Ndmas}$ be the set of  $\Ndmas \times \Ndmas$  diagonal matrices with positive diagonal entries. 
Specifically, we fix some $\epsilon > 0$ and restrict the diagonal entires of the matrices in  $\mySet{D}^{\Ndmas}$ to be not smaller than $\epsilon$.
We set the weights matrix $\myWeights$ to be the solution to:
\begin{equation}
\label{eqn:OptProb1}
\mathop{\min}\limits_{\myWeights \in \myWeightSet^{\Ndmas \times \Nantennas}, \tilde{\myMat{U}} \in \mySet{U}^{\Ndmas}, \tilde{\myMat{D}} \in \mySet{D}^{\Ndmas }} \left\|\myWeights -  \tilde{\myMat{U}} \tilde{\myMat{D}}\tilde{\myMat{V}}^H \CovW^{-1/2}\right\|^2.
\end{equation}

Let $P_{\myWeightSet}:\mySet{C}^{\Ndmas \times \Nantennas} \mapsto \myWeightSet^{\Ndmas \times \Nantennas}$ be the entry-wise projection into $\myWeightSet^{\Ndmas \times \Nantennas}$. 
By \eqref{eqn:Qmat1} the entry-wise projection of $\myMat{M} \in \mySet{C}^{\Ndmas \times \Nantennas}$ is given by
 	\ifsingle
\begin{equation*}
\left( P_{\myWeightSet}\left( \myMat{M}\right) \right)_{p_1, (p_2 -1)\Ndmas + l} = 
\begin{cases}
\mathop{\arg \min}\limits_{q \in \myWeightSet}\left|q - \left( \myMat{M}\right)_{p_1, (p_2 -1)\Ndmas + l}  \right|^2  & p_1 = p_2 \\
0 & p_1 \ne p_2.
\end{cases} 
 \end{equation*}
 \else
\begin{align*}
&\left( P_{\myWeightSet}\left( \myMat{M}\right) \right)_{p_1, (p_2 -1)\Ndmas + l} = \notag \\
&\qquad  
\begin{cases}
\mathop{\arg \min}\limits_{q \in \myWeightSet}\left|q - \left( \myMat{M}\right)_{p_1, (p_2 -1)\Ndmas + l}  \right|^2  & p_1 = p_2 \\
0 & p_1 \ne p_2.
\end{cases} 
\end{align*} 
 \fi 
 In order to solve \eqref{eqn:OptProb1}, we propose an alternating minimization algorithm, based on the properties detailed in the following lemma:
\begin{lemma}
	\label{lem:AltMin}
	For any $\myMat{M} \in \mySet{C}^{\Ndmas \times \Nantennas}$ we have that
	\begin{subequations}
		\label{eqn:AltMin}
	\begin{equation} 
	\label{eqn:AltMin1}
	\myWeights\am\left(\myMat{M}\right)  \triangleq  \mathop{\arg \min}\limits_{\myWeights \in \myWeightSet^{\Ndmas \times \Nantennas} } \left\|\myWeights - \myMat{M} \right\|^2 = P_{\myWeightSet}\left(\myMat{M} \right).
	\end{equation}
	Additionally, for any  $\myMat{M}_1, \myMat{M}_2 \in \mySet{C}^{\Ndmas \times \Nantennas}$, let $\myMat{U}_M$ and $\myMat{V}_M$ be the left singular vectors matrix and the right singular vectors matrix of $\myMat{M}_1\myMat{M}_2^H$, then 
 	\ifsingle
	\begin{equation} 
	\label{eqn:AltMin2}
	\tilde{\myMat{U}}\am\left(\myMat{M}_1, \myMat{M}_2 \right)  \triangleq  \mathop{\arg \min}\limits_{\tilde{\myMat{U}} \in \mySet{U}^{\Ndmas}} \left\|\myMat{M}_1 - \tilde{\myMat{U}}\myMat{M}_2 \right\|^2 
	=  \myMat{U}_M \myMat{V}_M^H.
	\end{equation}
	\else 
	\begin{align} 
	\tilde{\myMat{U}}\am\left(\myMat{M}_1, \myMat{M}_2 \right)  &\triangleq  \mathop{\arg \min}\limits_{\tilde{\myMat{U}} \in \mySet{U}^{\Ndmas}} \left\|\myMat{M}_1 - \tilde{\myMat{U}}\myMat{M}_2 \right\|^2 \notag \\
	&=  \myMat{U}_M \myMat{V}_M^H.
	\label{eqn:AltMin2}
	\end{align}	
	\fi 
	Finally, by letting $\myVec{m}_{1,i}$ and $\myVec{m}_{2,i}$ be the $i$th columns of $ \myMat{M}_1^H$ and $\myMat{M}_2^H$, respectively, $i \in \{1,2,\ldots,\Ndmas\}$, we have that the diagonal entries of   
	\begin{equation} 
	\label{eqn:AltMin3}
	\tilde{\myMat{D}}\am\left(\myMat{M}_1, \myMat{M}_2 \right)  \triangleq  \mathop{\arg \min}\limits_{\tilde{\myMat{D}} \in \mySet{D}^{\Ndmas}} \left\|\myMat{M}_1 - \tilde{\myMat{D}}\myMat{M}_2 \right\|^2,
	\end{equation}	
	are given by
	\begin{equation}
	\label{eqn:AltMin4}
	 \left( \tilde{\myMat{D}}\am\left(\myMat{M}_1, \myMat{M}_2 \right)\right)_{i,i} 
	 = \max \left(\frac{{\rm Re} \left(\myVec{m}_{1,i}^H\myVec{m}_{2,i}\right)} { \big\| \myVec{m}_{2,i}\big\|^2}, \epsilon  \right) . 
	\end{equation}	
	\end{subequations}
\end{lemma} 

{\em Proof:}
See Appendix \ref{app:Proof2}.

\smallskip
Based on Lemma \ref{lem:AltMin}, we propose to solve the joint optimization problem \eqref{eqn:OptProb1} in an alternating fashion, i.e., optimize over $\myWeights$ for   fixed $\tilde{\myMat{U}},\tilde{\myMat{D}}$, next optimize over  $\tilde{\myMat{U}}$ for   fixed   $\myWeights,\tilde{\myMat{D}}$, then optimize over $\tilde{\myMat{D}}$ for  fixed $\myWeights,\tilde{\myMat{U}}$,  and continue until convergence. The resulting alternating minimization algorithm is summarized in  Algorithm \ref{alg:Algo1}. 
As the Frobenious norm objective in \eqref{eqn:OptProb1} is differentiable, convergence of the alternating optimization algorithm is guaranteed \cite[Thm. 2]{Bezdek:03}. 

\begin{algorithm}
	\caption{\ac{dma} weights for identical frequency selectivity}
	\label{alg:Algo1}
	\begin{algorithmic}[1]
		\STATE Initialization: Set $k=0$ and $\tilde{\myMat{U}}_k = \myI_{\Ndmas}$, $\tilde{\myMat{D}}_k = \myI_{\Ndmas}$.
		\STATE \label{stp:MF0} Compute $\tilde{\myMat{V}}$ using Corollary \ref{cor:OptSetting}.
		\STATE \label{stp:MF1} Obtain  $\myWeights_{k\!+\!1}\!=\!\myWeights\am  $ with $\myMat{M} = \tilde{\myMat{U}}_k \tilde{\myMat{D}}_k\tilde{\myMat{V}}^H \CovW^{-1/2}$ using \eqref{eqn:AltMin1}.
		\STATE \label{stp:MF2} Set  $\tilde{\myMat{U}}_{k\!+\!1}\!=\! \tilde{\myMat{U}}\am  $ via \eqref{eqn:AltMin2} with $\myMat{M}_1 = \myWeights_{k \! + \! 1}$ and $\myMat{M}_2 =  \tilde{\myMat{D}}_k\tilde{\myMat{V}}^H \CovW^{-1/2} $.
		\STATE \label{stp:MF4} Set  $\tilde{\myMat{D}}_{k\!+\!1}\!=\! \tilde{\myMat{D}}\am  $ via \eqref{eqn:AltMin3}-\eqref{eqn:AltMin4} with $\myMat{M}_1 = \tilde{\myMat{U}}_{k\!+\!1}^H\myWeights_{k \! + \! 1}$ and $\myMat{M}_2 =   \tilde{\myMat{V}}^H \CovW^{-1/2} $. 
		\STATE If termination criterion is inactive: Set $k := k+1$ and go to Step \ref{stp:MF1}.	 
	\end{algorithmic}
\end{algorithm}

In Algorithm \ref{alg:Algo1} we exploit the fact that the optimal unconstrained $\myWeights\opt$ achieves the same sum-rate for any setting of $\tilde{\myMat{U}}, \tilde{\myMat{D}}$, and use these matrices as optimization variables. Consequently, we are able to obtain feasible weight matrices which are within a small distance from the optimal unconstrained matrix.  
In our numerical study in Section \ref{sec:Sims} we demonstrate that \acp{bs} equipped with \acp{dma} designed via Algorithm \ref{alg:Algo1} are capable of achieving performance which is within a small gap of the fundamental limits of the channel, achievable using optimal antenna arrays. Furthermore, it is illustrated that, unlike the unconstrained case, when $\myWeights \in \myWeightSet^{\Ndmas\times \Nantennas}$, increasing the number of microstrips $\Ndmas$ above $\Nusers$ increases the achievable sum-rate, as the resulting $\myWeights\opt$ can be better approximated using a feasible matrix.

\subsection{Practical Design for Arbitrary Frequency Selectivity}
\label{subsec:PracticalAFreq}
In the previous subsections we studied the special case where the wireless channel is frequency flat and the metamaterial elements exhibit the same frequency selectivity profile, i.e., $\MemG = 0$ and $\Hmat[\tau]=\myI_{n_t}\cdot h[\tau]$, for some scalar mapping $h[\tau]$. Under this setting, we were able to express the integral in \eqref{eqn:SumRate1} with a single log-det expression in \eqref{eqn:Case1SE}, as the wireless channel is memoryless and the effect of $h[\tau]$ on the transmitted signal was canceled by its contribution to the effective noise. However, wireless channels are typically frequency selective, and in practical metasurfaces, each element may exhibit a different frequency selectivity profile. Here, the frequency selectivity cannot be effectively canceled, and the  explicit value of $\Hmat[\tau]$ has to be accounted for. In the following we show how the design principles proposed in the previous subsections can be extended to this general case.

To that aim, we fix some positive integer $\Nint$, and define $\omega_i \triangleq \frac{2\pi \cdot i}{\Nint}$, $i \in \{1,2,\ldots,\Nint\}$. We can now approximate \eqref{eqn:SumRate1} as  
 	\ifsingle
\begin{align}
\Rs \!\approx\! \frac{1}{\Nint \cdot \Nusers}\mathop{\sum}\limits_{i=1}^{\Nint} \log \left|\myI_{\Ndmas}\! +\! \myWeights {\HmatFD} (\omega_i) \GmatFD(\omega_i)\GmatFD^H(\omega_i) {\HmatFD}^H(\omega_i)\myWeights^H \left(  \myWeights {\HmatFD}(\omega_i) \CovW{\HmatFD}^H(\omega_i)\myWeights^H  \right)^{-1}  \right|.
\label{eqn:SumRate2}
\end{align}
	\else
\begin{align}
\Rs \approx \frac{1}{\Nint \cdot \Nusers}\mathop{\sum}\limits_{i=1}^{\Nint} \log \bigg|&\myI_{\Ndmas} \!+\! \myWeights {\HmatFD} (\omega_i) \GmatFD(\omega_i)\GmatFD^H(\omega_i) {\HmatFD}^H(\omega_i)\myWeights^H \notag \\
&\left(  \myWeights {\HmatFD}(\omega_i) \CovW{\HmatFD}^H(\omega_i)\myWeights^H  \right)^{-1}  \bigg|.
\label{eqn:SumRate2}
\end{align}	
	\fi 
Note that as $\Nint$ increases, \eqref{eqn:SumRate2} approaches the actual sum-rate in \eqref{eqn:SumRate1}. 
\color{NewColor}
{We next write \eqref{eqn:SumRate2} in terms of a single log-det expression, as in \eqref{eqn:Case1SE}. To that aim, let $\BlkDiag\left(\{\myMat{A}_i\}_{i=1}^{\Nint} \right)$ be a block diagonal matrix with diagonal submatrices $\{\myMat{A}_i\}_{i=1}^{\Nint}$, and define the $\Nint\cdot \Nantennas \times \Nint\cdot \Nantennas$ block diagonal matrices $\bar{\Gmat} \triangleq \BlkDiag\left( \{ {\HmatFD} (\omega_i)\GmatFD(\omega_i)  \}_{i=1}^{\Nint}\right) $ and $\CovWBar \triangleq \BlkDiag\left( \{  {\HmatFD}(\omega_i) \CovW{\HmatFD}^H(\omega_i)  \}_{i=1}^{\Nint}\right) $. Also, define $\bar{\myWeights} \triangleq \myI_{\Nint} \otimes \myWeights$. Using these notations, it follows from  \cite[Pg. 122]{Meyer:00} that}
\ifsingle
\begin{align*}
&\myI_{\Nint \cdot \Ndmas}\! +\! \bar{\myWeights}   \bar{\Gmat}\bar{\Gmat}^H   \bar{\myWeights}^H \!\left(   \bar{\myWeights}  \CovWBar \bar{\myWeights}^H  \right)^{-1} \notag \\
&= \BlkDiag\left(\big\{\myI_{\Ndmas}\! +\! \myWeights {\HmatFD} (\omega_i) \GmatFD(\omega_i)\GmatFD^H(\omega_i) {\HmatFD}^H(\omega_i)\myWeights^H \left(  \myWeights {\HmatFD}(\omega_i) \CovW{\HmatFD}^H(\omega_i)\myWeights^H  \right)^{-1}\big\}_{i=1}^{\Nint} \right). 
\end{align*}
	\else
\begin{align*}
&\myI_{\Nint \cdot \Ndmas}\! +\! \bar{\myWeights}   \bar{\Gmat}\bar{\Gmat}^H   \bar{\myWeights}^H \!\left(   \bar{\myWeights}  \CovWBar \bar{\myWeights}^H  \right)^{-1} \notag \\
&= \BlkDiag\bigg(\big\{\myI_{\Ndmas}\! +\! \myWeights {\HmatFD} (\omega_i) \GmatFD(\omega_i)\GmatFD^H(\omega_i) {\HmatFD}^H(\omega_i)\myWeights^H \notag \\
&\qquad \qquad\qquad  \times  \left(  \myWeights {\HmatFD}(\omega_i) \CovW{\HmatFD}^H(\omega_i)\myWeights^H  \right)^{-1}\big\}_{i=1}^{\Nint} \bigg). 
\end{align*}	
	\fi 
Since $|\BlkDiag\left(\{\myMat{A}_i\}_{i=1}^{\Nint}\right) | = \prod_{i=1}^{\Nint}|\myMat{A}_i|$ when $\myMat{A}_i$ are square matrices \cite[Pg. 467]{Meyer:00}, it follows that \eqref{eqn:SumRate2} can be written as
\color{black}
\begin{align}
\Rs \!\approx\! \frac{1}{\Nint \!\cdot \Nusers}\! \log \left|\myI_{\Nint \cdot \Ndmas}\! +\! \bar{\myWeights}   \bar{\Gmat}\bar{\Gmat}^H   \bar{\myWeights}^H \!\left(   \bar{\myWeights}  \CovWBar \bar{\myWeights}^H  \right)^{-1}\!  \right|.
\label{eqn:Case2SE}
\end{align}

The approximation in \eqref{eqn:Case2SE} implies that the expression for the achievable sum-rate with arbitrary frequency selectivity is similar to that with identical frequency selectivity and flat channels in \eqref{eqn:Case1SE}. 
Consequently, the design principles used for configuring the \ac{dma} to minimize \eqref{eqn:Case1SE} in Algorithm~\ref{alg:Algo1} can also be used to minimize \eqref{eqn:Case2SE}. The main difference between minimizing \eqref{eqn:Case1SE} and \eqref{eqn:Case2SE} is that in \eqref{eqn:Case2SE}, the equivalent weights matrix $\bar{\myWeights}$ has to be written as $\myI_{\Nint} \otimes \myWeights$ where $\myWeights \in \myWeightSet^{\Ndmas \times \Nantennas}$. This additional constraint can be accounted for in the alternating minimization algorithm using the following lemma:
\begin{lemma}
	\label{lem:AidLemmaQ}
		For any $\myMat{M} \in \mySet{C}^{\Nint \cdot \Ndmas \times \Nint \cdot\Nantennas}$, the weights matrix which minimizes
			\begin{equation} 
			\label{eqn:AltMinQ1}
			\myWeights\ame\left(\myMat{M}\right)  \triangleq  \mathop{\arg \min}\limits_{\myWeights \in \myWeightSet^{\Ndmas \times \Nantennas} } \left\|\left( \myI_{\Nint} \otimes\myWeights\right)  - \myMat{M} \right\|^2,
			\end{equation} 
			is given by 
 	\ifsingle			
			\begin{equation} 
			\label{eqn:AltMinQ2}
			\left(\myWeights\ame\left(\myMat{M}\right)\right)_{p_1, (p_2 -1)\Ndmas + l} = 
			\begin{cases}
			\mathop{\arg \min}\limits_{q \in \myWeightSet}\sum\limits_{i=0}^{\Nint-1}  \left|q - \left( \myMat{M}\right)_{i \cdot\Ndmas + p_1, i \cdot \Nantennas + (p_2 -1)\Ndmas + l}  \right|^2  & p_1 = p_2 \\
			0 & p_1 \ne p_2.
			\end{cases} 
			\end{equation}
	\else
			\begin{align*} 
			&\left(\myWeights\ame\left(\myMat{M}\right)\right)_{p_1, (p_2 -1)\Ndmas + l} = \notag \\
			&\begin{cases}
			\mathop{\arg \min}\limits_{q \in \myWeightSet}\sum\limits_{i=0}^{\Nint-1}  \left|q - \left( \myMat{M}\right)_{i \cdot\Ndmas + p_1, i \cdot \Nantennas + (p_2 -1)\Ndmas + l}  \right|^2  & p_1 = p_2 \\
			0 & p_1 \ne p_2.
			\end{cases} 
			\end{align*}		
	\fi 
\end{lemma}

\begin{IEEEproof}
The lemma is obtained by explicitly writing the Frobenious norm in \eqref{eqn:AltMinQ1}, noting that each element of $\myWeights$ independently effects the overall norm via the sum of $\Nint$ terms, as given in the lemma. 
\end{IEEEproof}
 
Note that for $\myWeightSet = \mySet{C}$, the non-zero entries of $	\myWeights\ame$ are given by the sample mean of their corresponding entries in $\myMat{M}$.
Using Lemma \ref{lem:AidLemmaQ}, we can now adapt Algorithm \ref{alg:Algo1} to account for arbitrary frequency selectivity profiles, resulting in Algorithm~\ref{alg:Algo2}.

\begin{algorithm}
	\caption{\ac{dma} weights for arbitrary frequency selectivity}
	\label{alg:Algo2}
	\begin{algorithmic}[1]
		\STATE Initialization: Set $k=0$ and $\tilde{\myMat{U}}_k = \myI_{\Nint \cdot \Ndmas}$, $\tilde{\myMat{D}}_k = \myI_{\Nint \cdot\Ndmas}$.
		\STATE \label{stp:MFA0} Compute $\tilde{\myMat{V}}$ using Corollary \ref{cor:OptSetting} with $\CovW$ and $\Gmat$ replaced with $\CovWBar$ and $\bar{\Gmat}$, respectively.
		\STATE \label{stp:MFA1} Obtain  $\myWeights_{k\!+\!1}\!=\!\myWeights\ame  $ with $\myMat{M} = \tilde{\myMat{U}}_k \tilde{\myMat{D}}_k\tilde{\myMat{V}}^H \CovWBar^{-1/2}$ using \eqref{eqn:AltMin1}.
		\STATE \label{stp:MFA2} Set  $\tilde{\myMat{U}}_{k\!+\!1}\!=\! \tilde{\myMat{U}}\am  $ via \eqref{eqn:AltMin2} with $\myMat{M}_1 = \myI_{\Nint} \otimes \myWeights_{k \! + \! 1}$ and $\myMat{M}_2 =  \tilde{\myMat{D}}_k\tilde{\myMat{V}}^H \CovWBar^{-1/2} $.
		\STATE \label{stp:MFA4} Set  $\tilde{\myMat{D}}_{k\!+\!1}\!=\! \tilde{\myMat{D}}\am  $ via \eqref{eqn:AltMin3}-\eqref{eqn:AltMin4} with $\myMat{M}_1 = \tilde{\myMat{U}}_{k\!+\!1}^H\left( \myI_{\Nint} \otimes\myWeights_{k \! + \! 1}\right) $ and $\myMat{M}_2 =   \tilde{\myMat{V}}^H \CovWBar^{-1/2} $. 
		\STATE If termination criterion is inactive: Set $k := k+1$ and go to Step \ref{stp:MFA1}.	 
	\end{algorithmic}
\end{algorithm}

In order to evaluate the gap of the resulting configuration from optimality, we wish to characterize the maximal sum-rate achievable when $\myWeights$ can be any matrix in $\mySet{C}^{\Ndmas \times \Nantennas}$, not restricted to satisfy \eqref{eqn:Qmat1}. Since $\myWeights$ in \eqref{eqn:SumRate1} does not vary with $\omega$, obtaining the optimal performance is a difficult task. Nonetheless, \eqref{eqn:SumRate1} can be used to obtain an upper bound on the optimal average sum-rate, as stated in the following proposition:
\begin{proposition}
	\label{pro:ArbUpBound}
	If $\HmatFD(\omega)$ is non-singular for every $\omega \in [0, 2\pi)$, then the  maximal achievable average sum-rate is upper-bounded by
	\begin{equation}
	\label{eqn:ArbUpBound}
	\Rs \le  \frac{1}{2 \pi \cdot \Nusers}\mathop{\int}\limits_{0}^{2\pi}\sum\limits_{i=1}^{\min\left( \Nusers, \Ndmas\right) }\log\big(1+\lambda_i(\omega)\big) d \omega,
	\end{equation}  
	where $\{\lambda_i(\omega)\}_{i=1}^{\Nusers}$ are defined in Corollary \ref{cor:OptMIMO}.
\end{proposition}

{\em Proof:}
See Appendix \ref{app:Proof3}.
	
\smallskip	
\textcolor{NewColor}{It is emphasized that the upper bound in \eqref{eqn:ArbUpBound} is in general very difficult to approach in practice, as it is computed by allowing the \ac{dma} weights to be frequency selective, thus effectively canceling the frequency selectivity of the wireless channel and the different elements in the microstrips. As in our design we assume that the \ac{dma} weights do not vary in frequency, the resulting system in general cannot achieve the bound in Proposition \ref{pro:ArbUpBound}. 
	Nonetheless, in the numerical evaluations in Section \ref{sec:Sims} it is demonstrated that \acp{bs} utilizing practical \acp{dma} designed via Algorithm \ref{alg:Algo2} are capable of achieving performance which is comparable with the upper bound in \eqref{eqn:ArbUpBound}. In particular, we show that, when using properly configured \acp{dma}, the resulting achievable average sum-rate is within a reasonable gap from the upper bound and that both curves scale similarly with respect to \ac{snr}}
%
%

By repeating the arguments in the discussion following Corollary \ref{cor:OptSetting}, it holds that when $\Ndmas$ is not smaller than the rank of $\EqGMatFD(\omega)$ for each $\omega \in [0,2\pi]$, then the upper bound in \eqref{eqn:ArbUpBound} coincides with the fundamental performance limits $\Rs\Mopt$ given in Corollary \ref{cor:OptMIMO}.    

\section{Numerical Study}
\label{sec:Sims}
In this section we numerically evaluate the achievable performance using the \ac{dma} configurations derived in Section \ref{sec:SingleCSI}. 
First, in Subsection \ref{subsec:Sim_IdenFreq} we consider frequency flat channels with \acp{dma} in which each element exhibits the same frequency selectivity profile, and numerically evaluate the average sum-rate achievable using the \ac{dma} design in Algorithm \ref{alg:Algo1}. Then, in Subsection \ref{subsec:Sim_ArbFreq}, we study frequency selective channels with \acp{dma} in which each element exhibits a different frequency selectivity, and compute the achievable performance of the \ac{dma} configuration obtained using Algorithm \ref{alg:Algo2}.

\begin{figure}
		\centering
		\scalebox{0.48}{\includegraphics{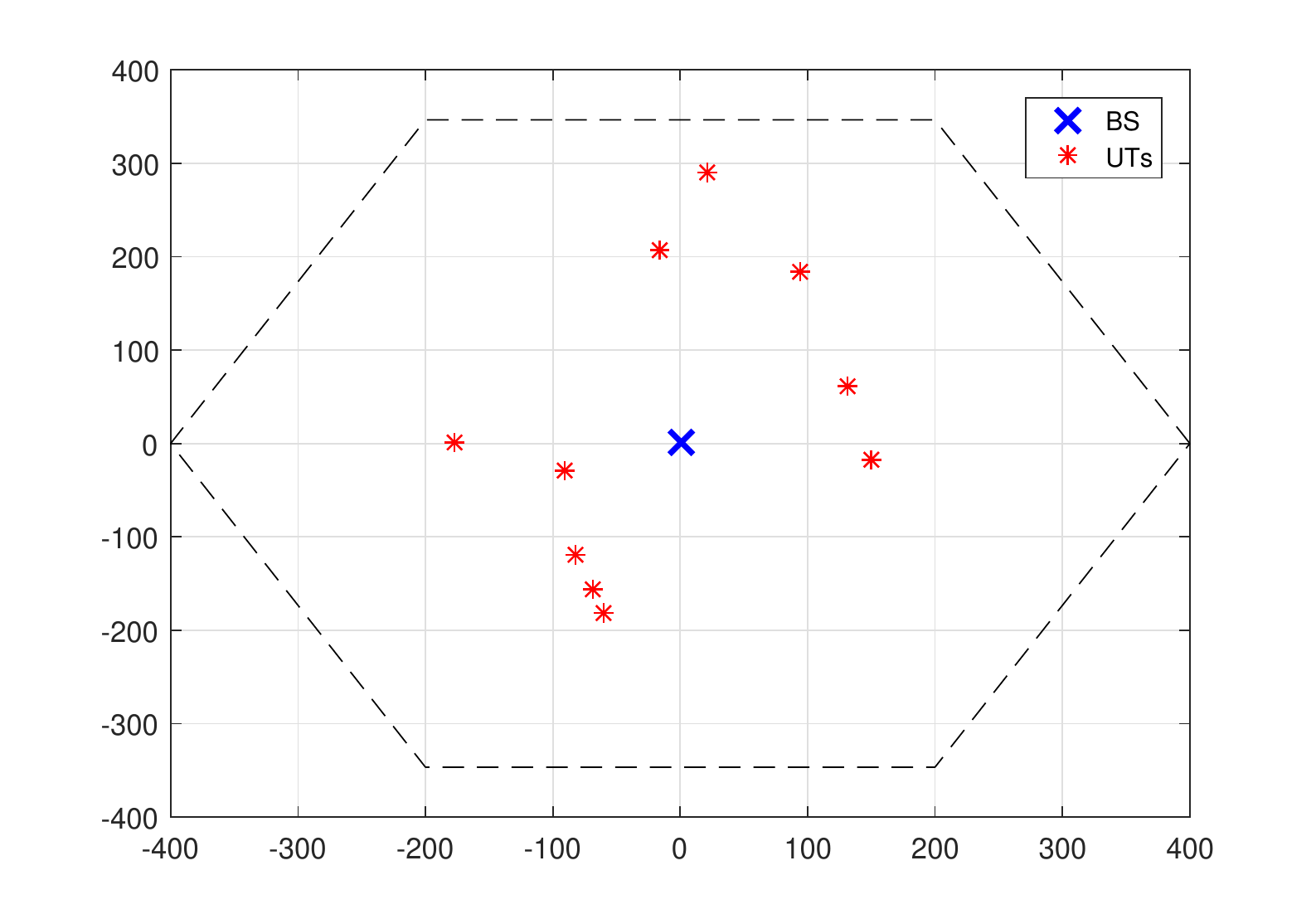}}
		\vspace{-0.4cm}
		\caption{Multi-user \ac{mimo} network illustration.}
		\label{fig:Network1}	
		\vspace{-0.2cm}	
	\end{figure}

We consider an uplink multi-user \ac{mimo} cell in a rich scattering environment. In this setup, a \ac{bs} equipped with a \ac{dma} serves $\Nusers = 10$ \acp{ut}, uniformly distributed in a hexagonal cell of radius $400$ m, with the exception of a circle with radius $20$ m around the \ac{bs}. An illustration of such a system is depicted in Fig. \ref{fig:Network1}. We use $\rho_i$ to denote the distance of the $i$th \ac{ut} from the \ac{bs}.  
Based on the model for frequency selective wireless \ac{mimo} channel proposed in \cite{Xiao:04}, the channel transfer matrices $\{\Gmat[\tau]\}_{\tau=0}^{\MemG}$ are generated as $\Gmat[\tau] =  \sigma_{\Gmat}^2[\tau] \CovMat{R}^{1/2} \Gmat_R[\tau]\myMat{D}[\tau]$, where:
\begin{itemize}
	\item $\{\sigma_{\Gmat}^2[\tau]\}_{\tau=0}^{\MemG}$ is the relative path loss of each tap, given by an exponentially decaying profile, i.e., $\sigma_{\Gmat}^2[\tau] = e^{-\tau}$.
	\item $\{\Gmat_R[\tau]\}_{\tau=0}^{\MemG}$ are a set of i.i.d. proper-complex zero-mean Gaussian ${\Nantennas\times \Nusers}$ matrices with i.i.d. entires of unit variance.
	\item $\CovMat{R}$ is an $\Nantennas \times \Nantennas$ representing the correlation induced by the sub-wavelength spacing of the elements in each microstrip. Neglecting the coupling between different microstrips, we set $\CovMat{R} = \myI_{\Ndmas} \otimes \CovMat{M}$, where $\CovMat{M} \in \mySet{C}^{\Nelements \times \Nelements}$ models the coupling induced between the elements of the same microstrip. In particular, we use Jakes' model\footnote{It is noted that Jakes' model requires  the radiating patterns to share the same azimuth \cite{Jakes:93}, which is a reasonable assumption for metasurface antenna elements.}  for the spatial correlation with element spacing of $0.2$ wavelength for $\CovMat{M}$, i.e.,  $\left( \CovMat{M}\right)_{i,l}  = J_0\big( 0.4\cdot\pi\cdot|i-l|\big) $, $i,l \in \{1,2,\ldots,\Nelements\}$, where $J_0(\cdot)$ is the zero-order Bessel function of the first type \cite{Jakes:93}. 
	\item $\{\myMat{D}[\tau] \}_{\tau=0}^{\MemG}$ are  $\Nusers\times \Nusers$ diagonal matrices representing the attenuation coefficients,  based on the model used in \cite{Marzetta:10}. In particular, we set $\left(\myMat{D}[\tau]\right)_{i,i} = \frac{\zeta_i[\tau]  }{\rho_i^2}$, where $\{\zeta_i[\tau] \}$ are the shadow fading coefficients, independently randomized from a log-normal distribution with standard deviation of $ 8$ dB. 
\end{itemize}


%
Since the radiating elements in the \ac{dma} microstrips are sub-wavelength separated, the additive noise $\myW[i]$ is inherently spatially correlated. Accounting for the coupling between the \ac{dma} elements, we set  $\CovW = \SigW \cdot \CovMat{R}$, \textcolor{NewColor}{where $\CovMat{R}$ is the matrix representing the correlation due to sub-wavelength element spacing defined above, and $\SigW >0$ models the average power of the noise signal.}

In the following we numerically evaluate the following achievable average sum-rates:
\begin{itemize}
	\item $\Rs^{\rm UC}$ - unconstrained weights, i.e., $\myWeightSet = \mySet{C}$. 
	\item $\Rs^{\rm AO}$ - amplitude only weights, here $\myWeightSet = [0.001,5]$.
	\item $\Rs^{\rm BA}$ - binary amplitude weights, $\myWeightSet = \{0, 0.1\}$.
	\item $\Rs^{\rm LP}$ - Lorentzian-constrained phase, namely, $\myWeightSet  =  \{\frac{j + e^{j \phi}}{2}: \phi \in [0, 2\pi]  \}$.
\end{itemize}
To compare the performance achievable with \acp{dma} to conventional analog combining, as in, e.g., \cite{Stein:17,AlKhateeb:14,Rial:16}, we also compute the rate when standard analog combining architectures with $\Ndmas$ RF chains are used. In particular, we simulate a fully connected phase shift network (Architecture A.1 in \cite{Rial:16}) and a fully connected switching network (Architecture A.3 in \cite{Rial:16}), both obtained using MaGiQ algorithm \cite[Sec. V-A]{Stein:17}. The resulting achievable average sum-rates are denoted $\Rs^{\rm AC}$ and $\Rs^{\rm SN}$, respectively. Since existing works on analog combining design assume memoryless channels\footnote{While it may be possible to extend analog combining design algorithm such as MaGiQ \cite{Stein:17} to frequency selective channels, such an extension is beyond the scope of this work.}, we simulate these setups only for the frequency flat scenarios in Subsection \ref{subsec:Sim_IdenFreq}.
 Our results are averaged over $1000$ Monte-Carlo simulations.   


\vspace{-0.2cm}
\subsection{Flat Channel with Identical Frequency Selectivity}
\label{subsec:Sim_IdenFreq}
\begin{figure}
		\centering
		\scalebox{0.48}{\includegraphics{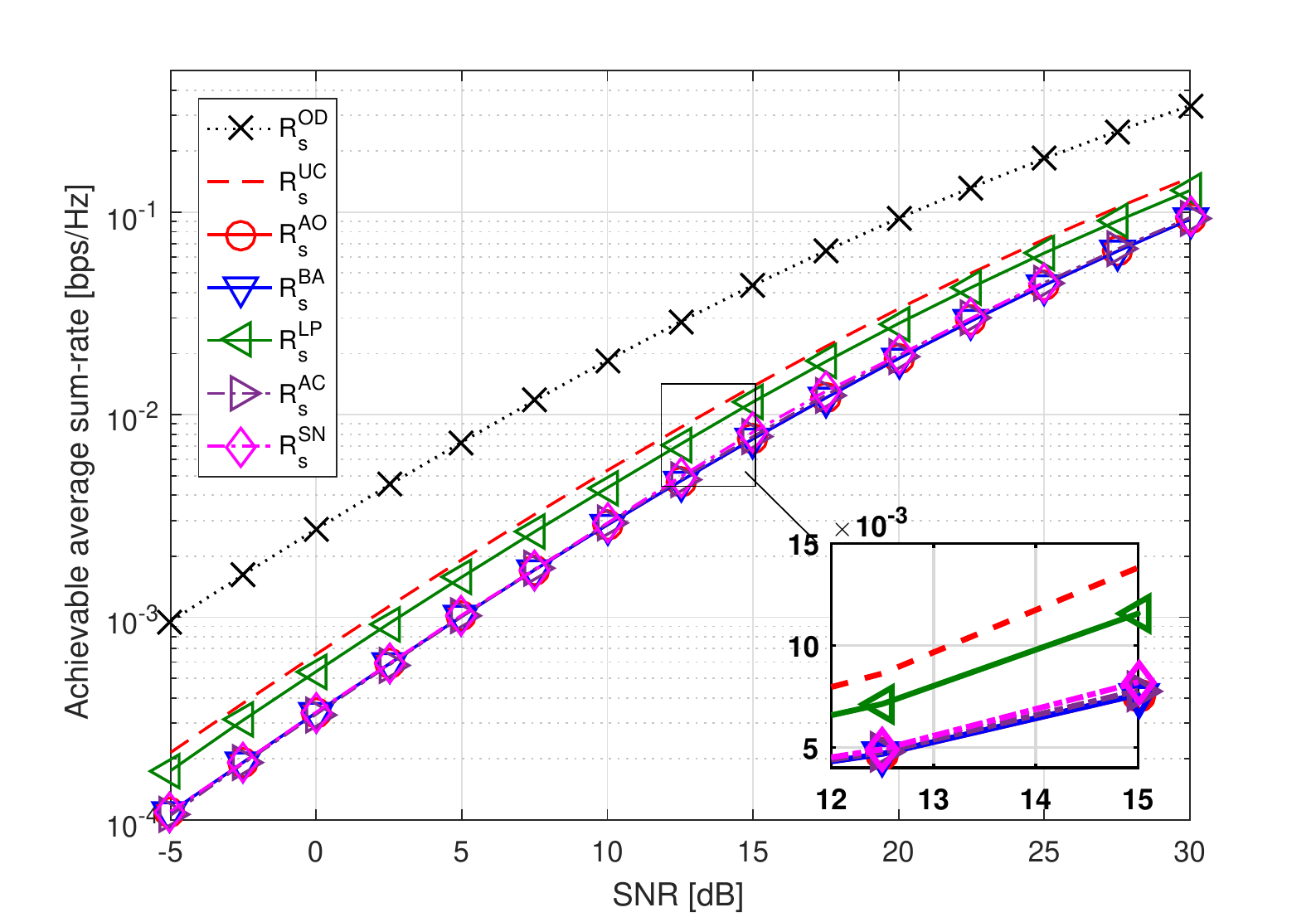}}
		\vspace{-0.8cm}
		\caption{Rate vs. SNR, flat channel, $\Nelements = 10$.}
		\label{fig:IFreq1}
		\vspace{-0.2cm}	
	\end{figure}

\begin{figure} 
		\centering
		\scalebox{0.48}{\includegraphics{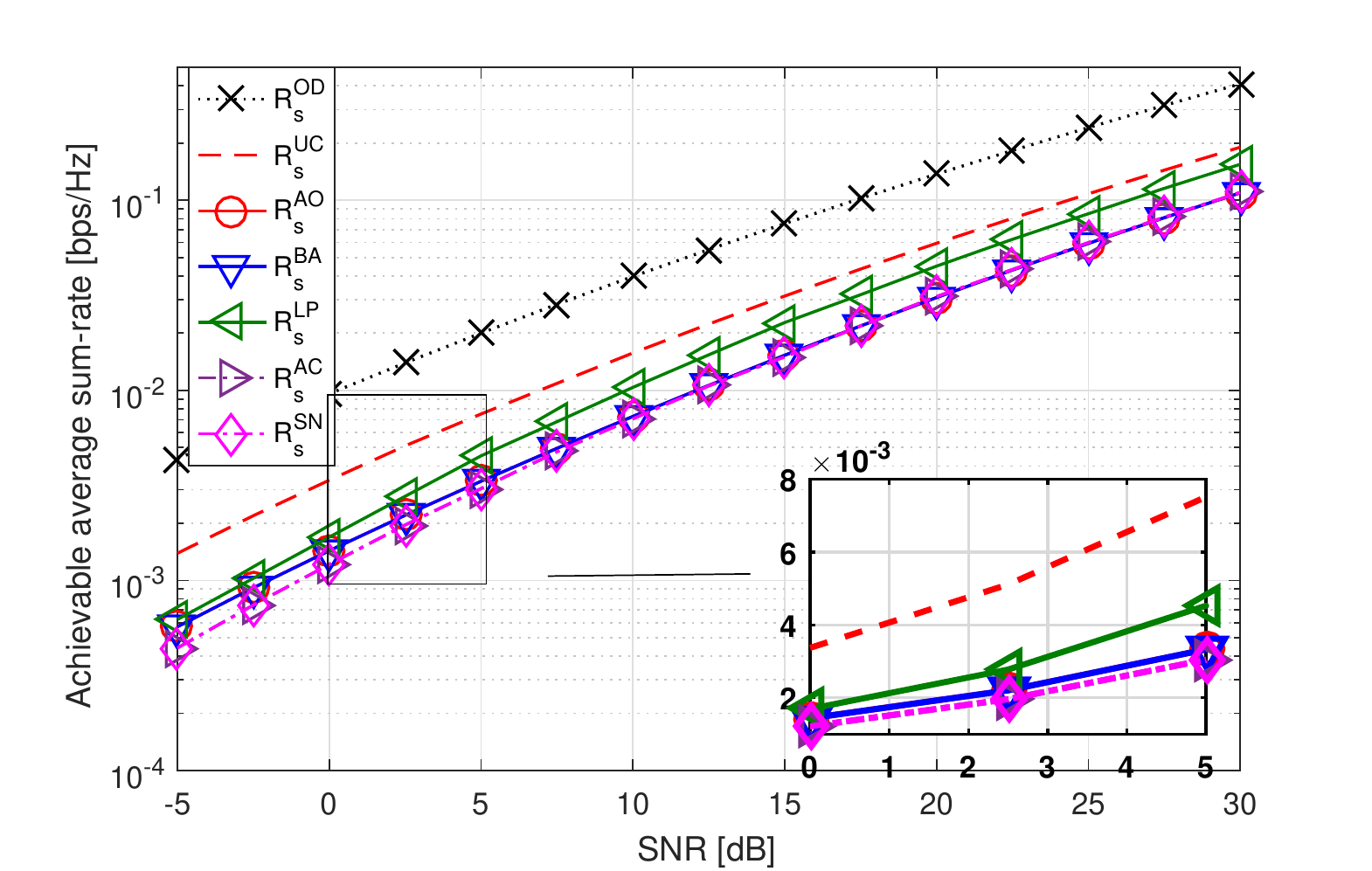}}
		\vspace{-0.2cm}
		\caption{Rate vs. SNR, flat channel, $\Nelements = 15$.}
		\label{fig:IFreq12}		
				\vspace{-0.2cm}
	\end{figure}

\vspace{-0.1cm}
We first consider the case where the channel is frequency-flat, namely, $\MemG = 0$, and each element in the \ac{dma} exhibits the same frequency selectivity profile, as studied in Subsections \ref{subsec:OptimalIFreq} and \ref{subsec:PracticalIFreq}. 
\label{txt:RateGap}
\textcolor{NewColor}{In Figs. \ref{fig:IFreq1}-\ref{fig:IFreq12} we let the \ac{snr}, defined as $1/\SigW$, vary in the range $[-5, 30]$ dB. Note that here the term \ac{snr} refers only the energy of the noise, and does not account for the attenuation induced by the channel, which depends on the specific realization of the location of each \ac{ut}. As the generated channels induce severe attenuation, the resulting achievable rate values are significantly smaller than those reported in previous related works, e.g., \cite{Rial:16, Mo:17}, in which the \ac{snr} encapsulates the channel attenuation. It is also noted that in the previous works  \cite{Rial:16, Mo:17} the rate measure represents the overall achievable rate in point-to-point \ac{mimo} communications, and not the achievable average sum-rate of a multi-user \ac{mimo} network, which can be viewed as the overall achievable rate divided by the number of \acp{ut}.}
  For each \ac{snr} value we compare the average sum-rates achievable using \acp{dma} with $\Ndmas = 10$ microstrips, each with $\Nelements$ radiating elements, computed via Algorithm \ref{alg:Algo1}, to the optimal achievable performance $\Rs\opt$ computed via Corollary \ref{cor:OptSetting}. In Fig. \ref{fig:IFreq1} we set $\Nelements = 10$ while in Fig. \ref{fig:IFreq12} we use $\Nelements = 15$.  Recall that since $\Ndmas \ge \Nusers$, then, based on the discussion following Corollary \ref{cor:OptSetting}, $\Rs\opt$ equals the fundamental performance limit, $\Rs\Mopt$, stated in Corollary \ref{cor:OptMIMO}. 
 Observing Figs. \ref{fig:IFreq1}-\ref{fig:IFreq12}, we note that $\Rs^{\rm LP}$ approaches $\Rs^{\rm UC}$ for all \ac{snr} values when $\Nelements =10$ and for \ac{snr} above $15$ dB when $\Nelements = 15$. This indicates that the Lorentzian-constrained phase restriction induces negligible loss when designing the weights using Algorithm \ref{alg:Algo1}. The amplitude only restriction, the binary amplitude constraint, the standard phase shifting network, and the standard switching network, all achieve roughly the same performance, which is within a small gap of that achievable using the Lorentzian-constrained phase weights. 
Furthermore, the \ac{snr} loss induced by restricting the weights matrix to satisfy \eqref{eqn:Qmat1}, namely, the fact that the \ac{dma} combines only inputs from the same microstrip, is approximately $7$ dB. In particular, for $\Nelements = 10$, $\Rs\opt$, which is achieved without the structure constraint \eqref{eqn:Qmat1}, achieves an average sum-rate of $0.1$ bps/Hz at \ac{snr} of $17$ dB, while $\Rs^{\rm UC}$ achieves the same performance for \ac{snr} of $24$ dB. For $\Nelements = 15$, $\Rs\opt = 0.1$ for \ac{snr} of $20$ dB, while $\Rs^{\rm UC}$ achieves this sum-rate at \ac{snr} of $27$ dB. 
Furthermore, it is noted that both curves scale similarly with respect to \ac{snr}, indicating that any average sum-rate which is achievable using an optimal unconstrained antenna array, is also achievable using practical \ac{dma} setups as the \ac{snr} increases.


\begin{figure} 
		\centering
		\scalebox{0.48}{\includegraphics{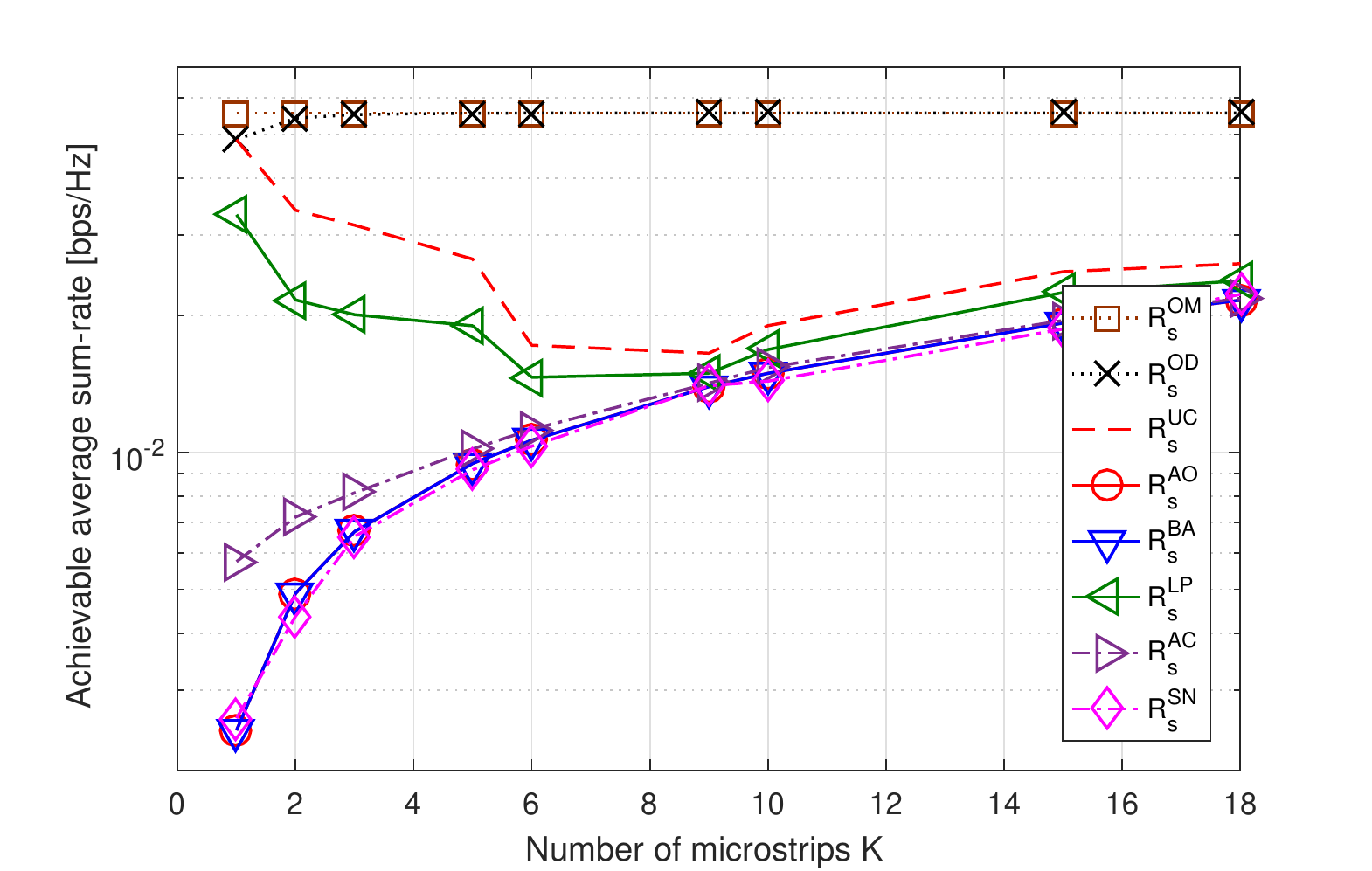}}
		\vspace{-0.2cm}
		\caption{Rate vs. microstrips, flat channel.}
		\label{fig:IFreq2} 
		\vspace{-0.2cm}
	\end{figure}

Next, in Fig. \ref{fig:IFreq2}, we fix the \ac{snr} to $15$ dB, the number of antennas to $\Nantennas = 90$, and compute the achievable average sum-rates for $\Ndmas \in [1,18]$.  
The goal of this study is to numerically evaluate how the number of microstrips effects the performance for a given number of antennas. 
In order to guarantee that the same channel and noise statistics are used for each value of $\Ndmas$, we fix the coupling matrix $\CovMat{R}$ to $\CovMat{R} = \myI_{15} \otimes \CovMat{M}$, where $\CovMat{M}$ is a $6\times 6$ matrix defined earlier in this section, representing the element coupling via Jakes' model. 
Observing Fig. \ref{fig:IFreq2}, we again note that the performance achievable with practical Lornetzian-constrained phase weights approaches that achieved with unconstrained weights for most considered values of $\Ndmas$, where the gap between $\Rs^{\rm LP}$ and the unconstrained $\Rs^{\rm UC}$ is at most $1.3\cdot 10^{-2}$ bps/Hz. 
\label{txt:SNRHit}
As expected, for $\Ndmas =1$, $\Rs^{\rm UC}$, which is subject only to \eqref{eqn:Qmat1}, coincides with the optimal performance $\Rs\opt$, as \eqref{eqn:Qmat1} imposes no constraint on the weights matrix structure  for $\Ndmas = 1$.  \textcolor{NewColor}{The performance gap of the \ac{dma}-based receivers from the optimal $\Rs\opt$ depends on the number of micropstrips $\Ndmas$. For example, for $\Ndmas = 6$, we observe that $\Rs^{\rm LP} = 1.5\cdot 10^{-2}$ bps/Hz, while  $\Rs^{\rm AO}$ and $\Rs^{\rm BA}$ are approximately $1\cdot 10^{-2}$ bps/Hz, i.e., gaps of roughly $4\cdot 10^{-2}$ bps/Hz  and $4.5\cdot 10^{-2}$ bps/Hz, respectively, from  the optimal performance $\Rs\opt = 5.5\cdot 10^{-2}$. This gap becomes less dominant as $\Ndmas$ further increases, and for $\Ndmas = 15$ it is reduced to approximately $3.5 \cdot 10^{-2}$ bps/Hz for all considered \ac{dma}-based receivers.}
Additionally, we note that $\Rs\opt$ is monotonically increasing for small values of $\Ndmas$, and for $\Ndmas > 3$ its value remains constant and equals the fundamental limit of the channel, $\Rs\Mopt$. This follows since, as noted in the discussion following Corollary \ref{cor:OptSetting}, $\Rs\opt$ remains constant when $\Ndmas$ is larger than the rank of $\tilde{\Gmat}$. 
\label{txt:Evals}
\textcolor{NewColor}{
Our numerical study shows that for the considered scenario, most realizations of $\tilde{\Gmat}$ have at most $3$ dominant eigenvalues. This behavior is due to the fact that the diagonal entries of the attenuation coefficients matrix $\myMat{D}$, which is randomized using the statistical model of \cite{Marzetta:10}, exhibit notable variations, as \acp{ut} located at different distances from the \ac{bs} can observe substantially different attenuation coefficients.
For this reason,} $\Rs\opt$ remains constant for $\Ndmas > 3$. 
Since  the constraint induced on $\Rs^{\rm UC}$ in \eqref{eqn:Qmat1} becomes less significant as $\Ndmas$ decreases, and since, as noted in Fig. \ref{fig:IFreq1}, $\Rs^{\rm LP}$ is capable of approaching $\Rs^{\rm UC}$ at such \acp{snr}, it is shown in Fig. \ref{fig:IFreq2} that, for a fixed number of elements $\Nantennas$, both $\Rs^{\rm UC}$ and $\Rs^{\rm LP}$ do not necessarily increase when the number of microstrips $\Ndmas$ is increased.  

While the results in Fig. \ref{fig:IFreq2} may be in favor of setting $\Ndmas = 1$ and $\Nelements = \Nantennas$, in practice increasing the number of elements on a single microstrip increases the attenuation which results from the propagation of the signal inside the microstrip. This phenomena is not accounted for in the model here, which assumes that the attenuation induced by each element is identical, hence the additional loss by increasing the number of elements per microstrip is not reflected in  Fig.~\ref{fig:IFreq2}. 
This observation also implies that in a practical implementation, we need to strike a balance between the cost (proportional to the number of RF ports or $\Ndmas$), losses (proportional to the number of metamaterial elements, $\Nelements$), and the performance (related to both quantities). This investigation is left for future works.

We also note that restricting the weights to binary values achieves roughly the same performance as continuous valued amplitude weights, and that under both constraints, the achievable average sum-rate substantially increases as $\Ndmas$ increases where the gap from the optimal \ac{mimo} $\Rs$ varies from approximately $6\cdot 10^{-2}$ bps/Hz for $\Ndmas = 1$ to $3.5\cdot 10^{-2}$ bps/Hz for $\Ndmas = 18$.  The standard analog combining networks $\Rs^{\rm AC}$ and $\Rs^{\rm SN}$, which also depend on $\Ndmas$ as its value here determines the number of RF chains, achieve approximately the same performance as \acp{dma} with continuous valued amplitude weights, where the phase shifters network achieves a slightly better performance for $\Ndmas < 9$.  
 
Finally, we observe in Fig. \ref{fig:IFreq2} that, while $\Rs\opt$ remains constant as the number of microstrips $\Ndmas$ increases above $\Nusers$, the performance achievable with \acp{dma} is monotonically increasing. This follows since, as discussed in Subsection \ref{subsec:PracticalIFreq}, increasing the number of microstrips $\Ndmas$ allows designing the matrices $\tilde{\myMat{U}}$ and $\tilde{\myMat{D}}$ in \eqref{eqn:OptimalQ}, which have no effect on the resulting optimal performance $\Rs\opt$, such that the resulting $\myWeights\opt$ can be better approximated using a feasible weights matrix. 

The results presented in this subsection demonstrate that for frequency flat channels, \acp{bs} equipped with \acp{dma} can achieve a performance which is comparable with costly optimal unconstrained antenna arrays. Furthermore, by utilizing our proposed alternating optimization algorithm, the sum-rate achievable with \acp{dma} is not smaller and even larger than that achieved using standard fully connected analog combiners, obtained using state-of-the-art design algorithms. 

\vspace{-0.2cm}
\subsection{Varying Frequency Selectivity}
\label{subsec:Sim_ArbFreq}
\vspace{-0.1cm}
Next, we consider the more general setup where the channel is frequency selective and each element exhibits a different frequency selectivity profile. In particular, we consider a channel with two taps, i.e., $\MemG = 1$, and set $\HmatFD(\omega) = \myI_{\Ndmas} \otimes  \HmatFD_G(\omega)$, where $\HmatFD_G(\omega) \in \mySet{C}^{\Nelements \times \Nelements }$ is a diagonal matrix representing the frequency selectivity profile of single microstrip. 
Based on the model detailed in Subsection \ref{subsec:Pre_DMA}, we account for the frequency response of the elements and the propagation inside the waveguide via the setting  $ \left( \HmatFD_G(\omega)\right)_{l,l} = e^{-(\alpha + j\cdot\beta (\omega) )\cdot l }$. 
In particular, we set $\beta (\omega) = 1.592\cdot \omega$ $[{\text m}^{-1}]$ and $\alpha = 0.0006$ $[{\text m}^{-1}]$, representing  a  microstrip with 50 ohm characteristic impedance made of Duroid 5880 operating at $1.9$ GHz with element spacing of $0.2$ wavelength (assuming free space wavelength) \cite[Ch. 3.8]{Pozar:09}. 



\begin{figure}
		\centering
		\scalebox{0.48}{\includegraphics{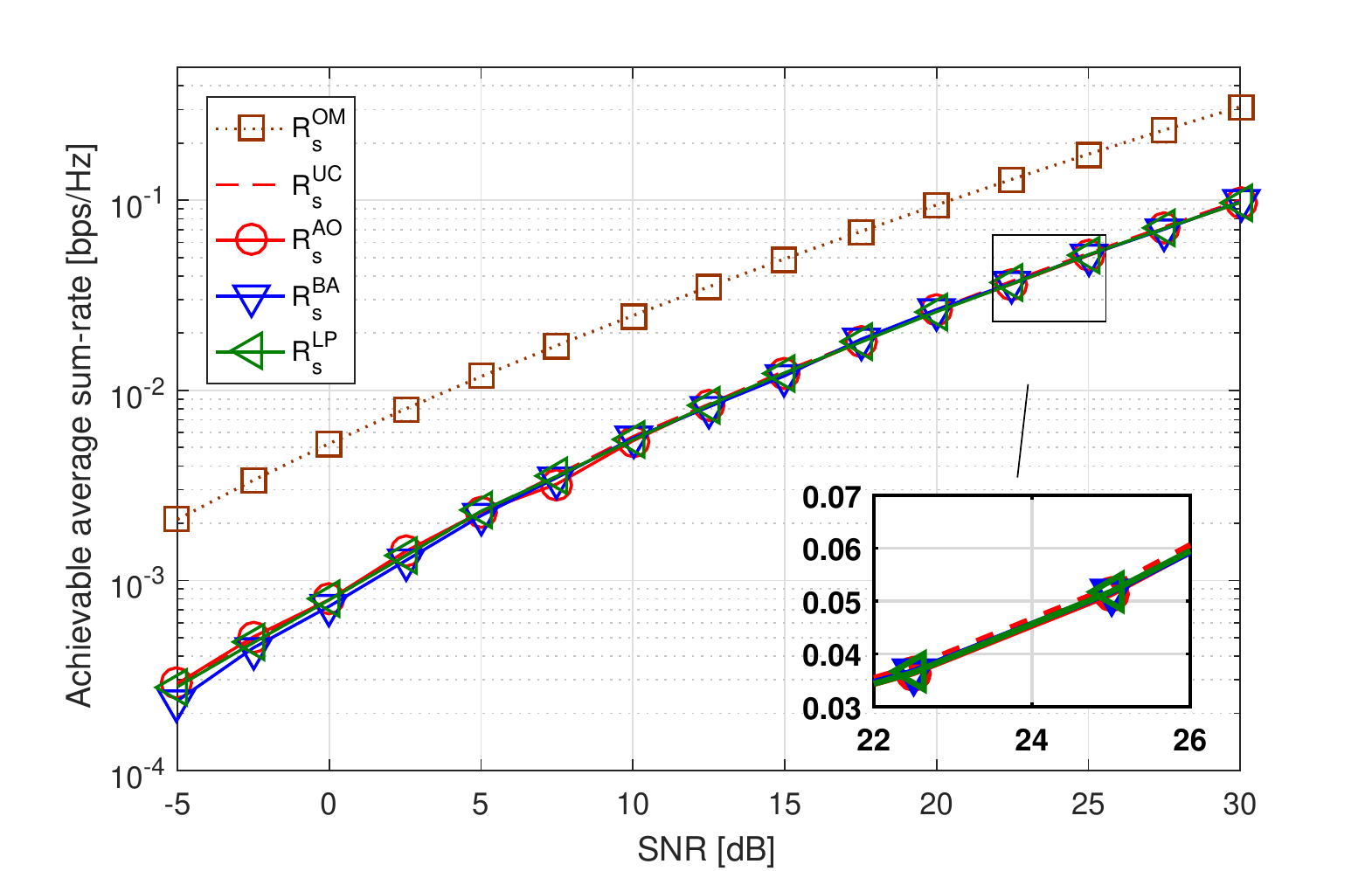}}
		\vspace{-0.2cm}
		\caption{Rate vs. SNR, frequency selective channel.}
		\label{fig:AFreq1}		
				\vspace{-0.2cm}
	\end{figure}

Fig. \ref{fig:AFreq1} depicts the average sum-rates achievable using \acp{dma} configured via Algorithm \ref{alg:Algo2} versus \ac{snr}, for $\Ndmas = 10$ and $\Nelements = 10$. The performance is compared to  the theoretical limit $\Rs\Mopt$.
Observing Fig. \ref{fig:AFreq1} we note that unlike the scenario considered in the previous subsection, here the performance achievable with \acp{dma} for all considered feasible sets $\myWeightSet$ is roughly the same. We also note that the achievable performance with \ac{dma} is within a notable gap of approximately $10$ dB in \ac{snr} from the upper bound on the maximal achievable performance in Proposition \ref{pro:ArbUpBound}.   The increased gap stems from the fact that, as shown in the proof of Proposition \ref{pro:ArbUpBound}, $\Rs\Mopt$ is obtained by mitigating the frequency selectivity of the wireless channel and the metamaterial elements. Unlike the scenario considered in Subsection \ref{subsec:Sim_IdenFreq} whose results are depicted in Fig. \ref{fig:IFreq1}, here the frequency selectivity induced by the physics of the metasurface cannot be mitigated by properly setting the coefficients matrix $\myWeights$, and thus the difference between upper bound $\Rs\Mopt$  and the achievable performance with \acp{dma} increases. Despite this gap, it is observed in Fig. \ref{fig:AFreq1} that the performance achievable with \acp{dma} scales similarly to the upper bound $\Rs\Mopt$ with respect to \ac{snr}, indicating that the performance achievable with \acp{dma} is comparable with $\Rs\Mopt$. 

	\begin{figure}
		\centering
		\scalebox{0.48}{\includegraphics{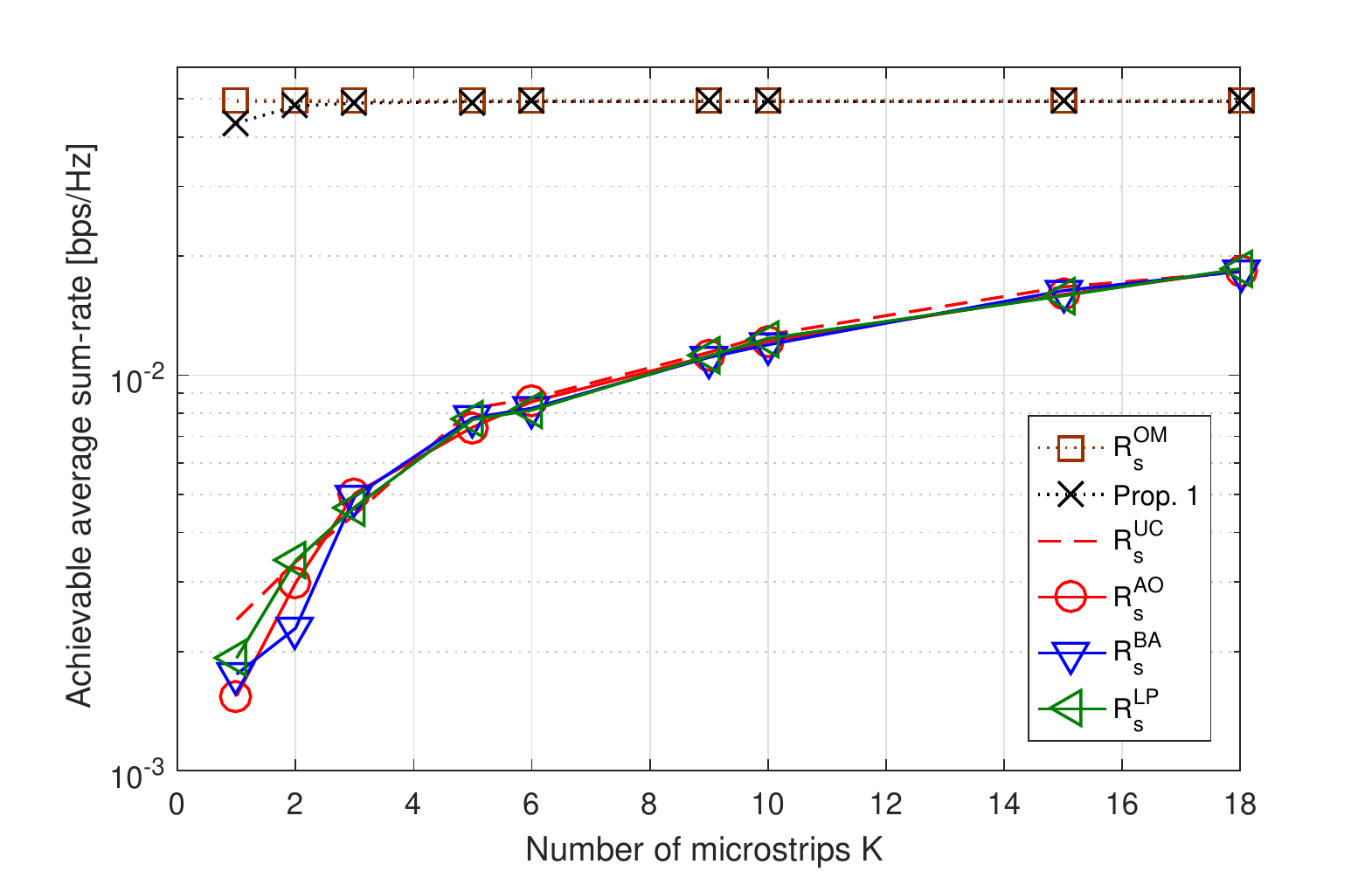}}
		\vspace{-0.2cm}
		\caption{Rate vs. microstrips, frequency selective channel.}
		\label{fig:AFreq2}
		\vspace{-0.2cm}		
	\end{figure}

Finally, in Fig. \ref{fig:AFreq2} we depict the achievable sum-rate versus number of microstrips for fixed number of antennas $\Nantennas = 90$ and \ac{snr} of $15$ dB. These rates are compared to the theoretical limit $\Rs\Mopt$, as well as to the upper bound on \ac{dma} performance computed via Proposition \ref{pro:ArbUpBound},    which coincides with the theoretical limit $\Rs\Mopt$ for $\Ndmas \ge \Nusers$.   
 Observing Fig. \ref{fig:AFreq2} we note that here, unlike the results depicted Fig. \ref{fig:IFreq2} which considered a similar scenario but did not account for the signal propagation inside the microsrtip, increasing the number of microstrips improves the performance. This follows since increasing the number of elements in each microsrtip induces additional attenuation, which impairs the ability of the \ac{bs} to recover the messages from the channel output. It is also observed in Fig. \ref{fig:AFreq2} that  the performance achievable using \acp{dma} is  roughly the same for all settings of $\myWeightSet$, as observed in Fig. \ref{fig:AFreq1}. The fact that the binary amplitude setting, which is relatively simple to implement, achieves roughly the same performance as the other settings, makes it an appealing candidate for future practical implementations studies. 
 Lastly, we note that there is a notable gap between the upper bound of Proposition \ref{pro:ArbUpBound}, which is computed by letting the \ac{dma} weights be frequency selective, and the actual sum-rate achievable with fixed weights. This indicates that the performance achievable with \acp{dma} can be substantially improved in frequency selective channels by designing the frequency response of each element, which is modeled in the \ac{dma} weights, to vary in frequency \textcolor{NewColor}{as in, e.g., \cite{Yoo:16}}. We leave the analysis and design of such \acp{dma} for future work.


\vspace{-0.2cm}
\section{Conclusions}
\label{sec:Conclusions}
\vspace{-0.1cm}
In this work we studied massive \ac{mimo} systems where the large-scale antenna array at the \ac{bs} is implemented using a \ac{dma}. We characterized the maximal achievable average sum-rate on the uplink, and derived two alternating optimization algorithms for designing practical \acp{dma} to approach the optimal performance: the first algorithm is designed for  frequency flat channels assuming that the frequency selectivity induced by the metasurface is identical among all elements, and the second algorithm generalizes the first algorithm to arbitrary multipath channels and frequency selectivity profiles. 
Our  results illustrate the potential gains over standard antenna arrays of utilizing \acp{dma} for implementing compact low-cost and low-power massive \ac{mimo} systems. In particular, it is shown in our simulations study that by properly adjusting the inherent combining and compression induced by the physics of \acp{dma}, a practical massive \ac{mimo} system can be constructed which is capable of achieving performance comparable to the fundamental theoretical limits, which require  costly, power consuming, and large-sized optimal antenna arrays.

\vspace{-0.2cm}
\begin{appendix}
\numberwithin{proposition}{subsection} 
\numberwithin{lemma}{subsection} 
\numberwithin{corollary}{subsection} 
\numberwithin{remark}{subsection} 
\numberwithin{equation}{subsection}	
%
\vspace{-0.2cm}
\subsection{Proof of Theorem \ref{thm:SumRate1}}
\label{app:Proof1}
\vspace{-0.1cm} 
To prove the theorem, we first write the operation of the \ac{dma} as part of the massive \ac{mimo} channel. By combining \eqref{eqn:Channel_Rel1} and \eqref{eqn:DMA_Rel1}, and defining $\tilde{\myW}[i] \triangleq \myWeights \sum\limits_{\tau=0}^{\Mem}\Hmat[\tau]\myW[i-\tau]$ and   $\tilde{\Hmat}[\tau] \triangleq \myWeights\sum\limits_{l=0}^{\MemG}\Hmat[\tau-l]\Gmat[l] $, the equivalent input-output relationship  can be written as 
\begin{equation}
\label{eqn:SingCellRel2}
\myZ[i]  =  \sum\limits_{\tau=0}^{\Mem + \MemG}\tilde{\Hmat}[\tau]\myX[i-\tau] + \tilde{\myW}[i].
\end{equation}

Note that $\tilde{\myW}[i]$ in \eqref{eqn:SingCellRel2} is a stationary multivariate proper-complex Gaussian process with finite memory $\Mem  + \MemG$. Consequently, \eqref{eqn:SingCellRel2} represents a finite-memory Gaussian \ac{mac}. Thus, by letting $\PSDMat{\myX}(\omega)$, $\PSDMat{\tilde{\myW}}(\omega)$,  and $\tilde{\HmatFD}(\omega)$, $\omega \in [0,2\pi)$, denote the \ac{psd} of $\myX[i]$, \ac{psd} of $\tilde{\myW}[i]$, and the \ac{dtft} of $\tilde{\Hmat}[\tau]$, respectively, it follows that the achievable average sum-rate is given by \cite{Verdu:89}
\ifsingle
\begin{align}
\!\!\Rs
&\!=\! \frac{1}{\Nusers}\mathop{\lim}\limits_{l \rightarrow \infty}\frac{1}{l}I\left(\myX^{l} ; \myZ^{l} \right) \notag \\
&\!=\! \frac{1}{2 \pi \cdot \Nusers}\mathop{\int}\limits_{0}^{2\pi} \log \left|\myI_{\Ndmas}\! +\! \tilde{\HmatFD}(\omega) \PSDMat{\myX}(\omega) \tilde{\HmatFD}^H(\omega) \left( \PSDMat{\tilde{\myW}}(\omega)\right)^{-1}  \right| d \omega \notag \\ 
&\!\stackrel{(a)}{=}\! \frac{1}{2 \pi \cdot \Nusers}\mathop{\int}\limits_{0}^{2\pi} \log \bigg|\myI_{\Ndmas}\! + \!\myWeights {\HmatFD}(\omega) \GmatFD(\omega)\GmatFD^H(\omega) {\HmatFD}^H(\omega)\myWeights^H    \left(  \myWeights {\HmatFD}(\omega) \CovW{\HmatFD}^H(\omega)\myWeights^H  \right)^{-1}  \bigg| d \omega,
\label{eqn:SumRate1a}
\end{align}
\else
\begin{align}
\!\!\Rs
&\!=\! \frac{1}{\Nusers}\mathop{\lim}\limits_{l \rightarrow \infty}\frac{1}{l}I\left(\myX^{l} ; \myZ^{l} \right) \notag \\
&\!=\! \frac{1}{2 \pi \cdot \Nusers}\mathop{\int}\limits_{0}^{2\pi} \log \left|\myI_{\Ndmas}\! +\! \tilde{\HmatFD}(\omega) \PSDMat{\myX}(\omega) \tilde{\HmatFD}^H(\omega) \left( \PSDMat{\tilde{\myW}}(\omega)\right)^{-1}  \right| d \omega \notag \\ 
&\!\stackrel{(a)}{=}\! \frac{1}{2 \pi \cdot \Nusers}\mathop{\int}\limits_{0}^{2\pi} \log \bigg|\myI_{\Ndmas}\! + \!\myWeights {\HmatFD}(\omega) \GmatFD(\omega)\GmatFD^H(\omega) {\HmatFD}^H(\omega)\myWeights^H \notag \\
&\qquad \qquad \qquad  \times  \left(  \myWeights {\HmatFD}(\omega) \CovW{\HmatFD}^H(\omega)\myWeights^H  \right)^{-1}  \bigg| d \omega,
\label{eqn:SumRate1a}
\end{align}
\fi 
where $(a)$ follows since $\myX[i]$ and $\myW[i]$ are i.i.d. processes with covariances $\myI_{\Nusers}$ and $\CovW$, respectively. Note that \eqref{eqn:SumRate1a} coincides with \eqref{eqn:SumRate1}, thus proving the theorem.

\vspace{-0.2cm}
\subsection{Proof of Lemma \ref{lem:AltMin}}
\label{app:Proof2}
\vspace{-0.1cm} 
	The equality \eqref{eqn:AltMin1} follows from the definition of the Frobenious norm, i,e,. 
	\begin{equation*}
	\left\|\myWeights - \myMat{M} \right\|^2 = \sum\limits_{i=1}^{\Ndmas}\sum\limits_{l=1}^{\Nantennas} \left|\left(\myWeights \right)_{i,l} - \left(\myMat{M} \right)_{i,l} \right|^2.
	\end{equation*}
	Since the feasible set $\myWeightSet^{\Ndmas \times \Nantennas}$ is defined entry-wise,   the Frobenious norm is minimized by entry-wise projection.  
	
	Similarly, the minimizing diagonal matrix in \eqref{eqn:AltMin3}-\eqref{eqn:AltMin4} is obtained since 
	\ifsingle	
	\begin{align*}
	\left\|\myMat{M}_1 - \tilde{\myMat{D}}\myMat{M}_2 \right\|^2 
	&= \sum\limits_{i=1}^{\Ndmas} 	\left\|\myVec{m}_{1,i} - \big( \tilde{\myMat{D}}\big)_{i,i} \cdot \myVec{m}_{2,i} \right\|^2 \notag \\
	&\stackrel{(a)}{=} \sum\limits_{i=1}^{\Ndmas} \big\| \myVec{m}_{1,i} \big\|^2 -2 {\rm Re} \left(\myVec{m}_{1,i}^H\myVec{m}_{2,i}\right) \cdot \big( \tilde{\myMat{D}}\big)_{i,i} + \big\| \myVec{m}_{2,i} \big\|^2 \cdot \big( \tilde{\myMat{D}}\big)_{i,i}^2,
	\end{align*}
	\else 
	\begin{align*}
	&\left\|\myMat{M}_1 - \tilde{\myMat{D}}\myMat{M}_2 \right\|^2 
	= \sum\limits_{i=1}^{\Ndmas} 	\left\|\myVec{m}_{1,i} - \big( \tilde{\myMat{D}}\big)_{i,i} \cdot \myVec{m}_{2,i} \right\|^2 \notag \\
	&\stackrel{(a)}{=} \sum\limits_{i=1}^{\Ndmas} \big\| \myVec{m}_{1,i} \big\|^2\! -\! 2 {\rm Re} \left(\myVec{m}_{1,i}^H\myVec{m}_{2,i}\right) \cdot \big( \tilde{\myMat{D}}\big)_{i,i}\! + \! \big\| \myVec{m}_{2,i} \big\|^2 \! \cdot \big( \tilde{\myMat{D}}\big)_{i,i}^2,
	\end{align*}	
	\fi 
	where $(a)$ holds as $\big( \tilde{\myMat{D}}\big)_{i,i}$ is real-valued. 
	Consequently, for each $i \in \{1,2,\ldots,\Ndmas\}$, the optimal setting of $\big( \tilde{\myMat{D}}\big)_{i,i} \ge \epsilon$ is given by \eqref{eqn:AltMin4}.
	
	Finally, the minimizing unitary matrix in \eqref{eqn:AltMin2} is obtained from the unitary Procrustes problem  \cite[Ch. 7.4]{Horn:90}, concluding the proof of the lemma.

\vspace{-0.2cm}
\subsection{Proof of Proposition \ref{pro:ArbUpBound}}
\label{app:Proof3}
\vspace{-0.1cm} 	
	The proposition is obtained by letting $\myWeights$ vary with $\omega$, i.e., replacing $\myWeights$ in \eqref{eqn:SumRate1} with $\myWeights(\omega)$. 
	Next, we  define $\tilde{\myWeights}(\omega) \triangleq  \myWeights(\omega)\HmatFD(\omega)$. Since  $\HmatFD(\omega)$ is non-singular, $\myWeights(\omega)$ can be recovered from $\tilde{\myWeights}(\omega)$. Under this setting, \eqref{eqn:SumRate1} satisfies
	\ifsingle	
	\begin{align}
	\Rs &= \frac{1}{2 \pi }\mathop{\int}\limits_{0}^{2\pi} \frac{1}{\Nusers}\log \bigg| \myI_{\Ndmas} +  \myWeights (\omega) \GmatFD(\omega)\GmatFD^H(\omega)  \myWeights^H(\omega)   \left(  \myWeights(\omega)  \CovW \myWeights^H(\omega)  \right)^{-1}\!  \bigg| d \omega \notag \\
	&\stackrel{(a)}{\le}  \frac{1}{2 \pi }\mathop{\int}\limits_{0}^{2\pi} \Rs\opt = \Rs\opt,
	\label{eqn:ProofProp1}
	\end{align}  
	\else
	\begin{align}
	\Rs &= \frac{1}{2 \pi }\mathop{\int}\limits_{0}^{2\pi} \frac{1}{\Nusers}\log \bigg| \myI_{\Ndmas} +  \myWeights (\omega) \GmatFD(\omega)\GmatFD^H(\omega)  \myWeights^H(\omega) \notag \\
	&\qquad \qquad \qquad \qquad \qquad \times  \left(  \myWeights(\omega)  \CovW \myWeights^H(\omega)  \right)^{-1}\!  \bigg| d \omega \notag \\
	&\stackrel{(a)}{\le}   \frac{1}{2 \pi \cdot \Nusers}\mathop{\int}\limits_{0}^{2\pi}\sum\limits_{i=1}^{\min\left( \Nusers, \Ndmas\right) }\log\big(1+\lambda_i(\omega)\big) d \omega,
	\label{eqn:ProofProp1}
	\end{align} 
	\fi 
	where $(a)$ follows from upper-bounding the integrand for each $\omega \in [0,2\pi)$, using Corollary \ref{cor:OptSetting}, thus proving  \eqref{eqn:ArbUpBound}. 

\end{appendix}	

\end{document}